\documentclass[showpacs,aps,prd,reprint,groupedaddress,nofootinbib]{revtex4-1}

\usepackage{amsmath,amssymb}
\usepackage{bm}
\usepackage{graphicx}
\usepackage{ascmac}
\usepackage{color}
\usepackage{longtable}
\usepackage{mathrsfs}

\usepackage{slashed}

\allowdisplaybreaks[1]

\usepackage[normalem]{ulem}  

\allowdisplaybreaks[3]

\begin{document}


\title{Kondo effect in charm/bottom nuclei}


\author{Shigehiro~Yasui}
\email[]{yasuis@th.phys.titech.ac.jp}
\affiliation{Department of Physics, Tokyo Institute of Technology, Tokyo 152-8551, Japan}



\begin{abstract}
We discuss the Kondo effect for isospin-exchange interaction between a $\bar{D}$, $B$ meson and a valence nucleon in charm/bottom atomic nuclei including the discrete energy-levels for valence nucleons.
To investigate the binding energy by the Kondo effect, we introduce the mean-field approach for the bound state of the $\bar{D}$, $B$ meson in charm/bottom nuclei.
Assuming a simple model, we examine the validity of the mean-field approximation by comparing the results with the exact solutions.
We also estimate the effect of the quantum fluctuation beyond the mean-field approximation.
We discuss the competition between the Kondo effect and the other correlations in valence nucleon, the isospin symmetry breaking and the  nucleon pairings.
\end{abstract}

\pacs{12.39.Hg,14.40.Lb,14.40.Nd,21.85.+d}
\keywords{Charm/bottom nuclei, Kondo effect, Heavy quark effective theory}

\maketitle


\section{Introduction}

In recent hadron and nuclear physics, the flavors of nuclear systems are extended to multi-flavor direction; strangeness flavor for hypernuclei and $\bar{K}$-mesic nuclei.
In these days, nuclear systems with charm/bottom flavor (charm/bottom nuclei) are also investigated in many theoretical studies.
One of the most interesting properties in charm/bottom nuclei is that the masses of charm/bottom hadrons are much heavier than the nucleon mass.
For example, the mass of the lightest charm meson, a $\bar{D}$ meson, is 1870 MeV, which is about twice as large as the nucleon mass.
The mass of the lightest bottom meson, a $B$ meson, is 5280 MeV, which is even about 5.6 times as large as the nucleon mass.
To investigate the behavior of such a heavy particle in nuclear system is an interesting problem as the impurity physics.
This is important, not only for understanding the hadron dynamics (hadron interaction, change of hadron in medium) and the nuclear structure, but also for  unveiling the fundamental properties of the strong interaction, such as the spontaneous breaking of chiral symmetry and the color confinement in vacuum, in Quantum Chromodynamics (QCD).
In fact, it is discussed that the ``heavy-quark spin symmetry" as the fundamental symmetry in QCD is essentially important in the heavy hadron interaction and the mass spectroscopy of heavy hadrons \cite{Neubert:1993mb,Manohar:2000dt,Yasui:2013vca,Yamaguchi:2014era}.

Existence of impurities can affect the properties of the matter systems.
As the famous and important impurity physics in condensed matter systems, the Kondo effect has been investigated for a long time \cite{Kondo:1964,Hewson,Yamada}.
We consider that the valence fermion (the quasi-particle forming the Fermi surface in medium; electrons in metal) and the impurity particle (atom with finite spin) has the spin-dependent force ($\vec{s} \cdot \vec{S}$ type interaction with spin operators $\vec{s}$ and $\vec{S}$ for the valence fermion and the impurity particle, respectively) with SU(2) spin symmetry.
Then, the effective interaction in low-energy scattering becomes enhanced due to the non-cancellation of loop effects from multiple number of particle-hole pairs around the Fermi surface, and the system becomes a strongly coupled one regardless to the small coupling between the valence fermion and the impurity particle.
As results, the impurity particle with spin-exchange interaction can change the transport properties of the condensed matter systems.
This is called the Kondo effect.
The conditions for emergence of the Kondo effect are: (i) heavy impurity, (ii) existence of Fermi surface (or degenerate state), (iii) loop contribution as quantum effect and (iv) non-Abelian interaction, such as the spin-dependent force \cite{Yamada}.
As far as those conditions are satisfied, the Kondo effect can emerge in any quantum systems for any kind of constituent particle and energy scale.

Recently, the Kondo effect is discussed for the isospin-exchange interaction with SU(2) isospin symmetry between a charm/bottom hadron and a nucleon in nuclear matter, and for the color-exchange interaction with SU(3) color symmetry between a light (up, down, strange) quark and a heavy (charm, bottom) quark \cite{Yasui:2013xr,Hattori:2015hka}.
The Kondo effect in strong magnetic fields is also discussed \cite{Ozaki:2015sya}.
Those can be studied in experimental studies in high energy accelerator facilities.
To research the Kondo effect in charm/bottom atomic nuclei, it is important to consider finite-volume effects and discrete energy levels of valence nucleons. 
Here we mean that a valence nucleon is the nucleon which is an active degree of freedom in a model space in shell-structure in atomic nuclei.
In the present work, focusing on the discrete energy-levels of valence nucleons, we study the Kondo effect in charm/bottom nuclei.

As mentioned, the non-Abelian interaction is one of the essential conditions for the Kondo effect.
In general, it is known that there are several kinds of non-Abelian interaction in (charm/bottom) nuclei;
(i) interaction changing  total angular momentum (sum of spin and orbital angular momentum) of valence nucleon,
(ii) interaction changing heavy-quark spin, and 
(iii) interaction changing isospin of heavy hadron and valence nucleon.

(i) The first induces the Kondo effect in deformed nuclei whose shape is different from the spherical one.
This may be a phenomena irrelevant to heavy impurity.
Instead, the coupling of nucleon spin to quantum rotation of the deformed nucleus is important.
In Ref.~\cite{Sugawara-Tanabe:1979}, Sugawara-Tanabe and Tanabe argued that the Coriolis force in deformed nucleus plays the interesting role of non-Abelian interaction.
In this case, the Coriolis force compels the spins of valence nucleons aligned along the spin of the deformed nuclei in the same direction (anti-pairing force), and hence the Kondo effect reduces the strength of the effective coupling 
 in the low-energy limit.

(ii) The second is the interaction in the heavy-quark effective theory based on QCD, which is given by $1/m_{\mathrm{Q}}$ expansion for the heavy quark mass $m_{\mathrm{Q}}$ \cite{Neubert:1993mb,Manohar:2000dt}.
It is known that the heavy-quark spin is the conserved quantity in the heavy-quark limit ($m_{\mathrm{Q}} \rightarrow \infty$), regardless to the non-perturbative interaction to light quarks and gluons. 
In charm/bottom nuclei with a bound $\Lambda_{\mathrm{c}}$, $\Lambda_{\mathrm{b}}$ baryon (isospin $1/2$, spin-parity $1/2^+$) \cite{Liu:2011xc,Maeda:2015hxa}, the heavy-quark spin is carried by the $\Lambda_{\mathrm{c}}$, $\Lambda_{\mathrm{b}}$ baryon, when only the leading order in the $1/m_{\mathrm{Q}}$ expansion is considered.
In the heavy-quark limit, the heavy-quark spin, namely the spin of $\Lambda_{\mathrm{c}}$, $\Lambda_{\mathrm{b}}$ baryon, cannot flip, and hence cannot induce the Kondo effect in charm/bottom nuclei.

(iii) The third gives the Kondo effect by isospin exchange between a $\bar{D}$, $B$ meson (isospin $1/2$, spin-parity $0^-$) and a valence nucleon. The isospin exchange is still available in the heavy-quark limit, because the isospin-degrees of freedom remain for the $\bar{D}$, $B$ meson in this limit.
It has been discussed by many theoretical studies that the $\bar{D}$, $B$ meson can be bound in nuclear matter\footnote{The dynamics of a $\bar{D}$, $B$ meson ($q\bar{Q}$; $Q=c$, $b$) is simpler than that of the antiparticle, a $D$, $\bar{B}$ meson ($\bar{q}Q$). Because the former does not have $q\bar{q}$ annihilation in nuclear matter, while the latter has. The difference of their behaviors is due to the charge symmetry breaking at finite baryon number density.}; the quark-meson coupling model \cite{Tsushima:1998ru,Sibirtsev:1999js}, the QCD sum rules \cite{Hayashigaki:2000es,Hilger:2008jg,Azizi:2014bba,Suzuki:2015est}, the hadron-interaction model \cite{Mishra:2003se,Mishra:2008cd,Kumar:2010gb,Kumar:2011ff,Lutz:2005vx,Tolos:2007vh,JimenezTejero:2011fc,GarciaRecio:2011xt,Yasui:2012rw,Yasui:2013iga,Suenaga:2014dia,Suenaga:2015daa}, and the quark-interaction model \cite{Blaschke:2011yv}.
Some of them suggest that a $\bar{D}$, $B$ meson is bound by attractive potential in nuclear matter.
It is interesting that the pion-exchange interaction between a $\bar{D}$, $B$ meson and a nucleon can be attractive enough to form some bound/resonant states \cite{Cohen:2005bx,Yasui:2009bz,Yamaguchi:2011xb,Yamaguchi:2011qw,Yamaguchi:2013hsa}\footnote{The bound/resonant systems composed of a $\bar{D}$, $B$ meson and a nucleon were originally investigated by the bound-state approach in the Skyrme model as pentaquark states \cite{Riska:1992qd,Oh:1994np,Oh:1994yv}.}.
In the present study, we investigate the $\bar{D}$, $B$ meson as the impurity for the Kondo effect in charm/bottom nuclei.
We note that a $\Lambda_{\mathrm{c}}$, $\Lambda_{\mathrm{b}}$ has no isospin and hence does not induce Kondo effect by isospin-exchange interaction.

Theoretically, to obtain correctly the ground state of the system with the Kondo effect is a highly non-perturbative problem, because the system becomes a strongly-coupled one due to the enhancement of the coupling strength in the low-energy scattering.
Several theoretical approaches have been developed for this problem \cite{Hewson,Yamada}.
One of the most effective methods is the numerical renormalization group initiated by Wilson \cite{RevModPhys.47.773}.
In the present analysis, we will adopt the mean-field approach \cite{Takano:1966,Yoshimori:1970}.
This has been applied also to the quantum dot systems with the Kondo effect \cite{Eto:2001}.
The mean-field approach provides us with a useful method for theoretical analysis and gives an intuitive understanding of the Kondo effect.
We recall that the idea of the mean-field approximation, or the Hartree-Fock approximation, has been known to be very useful in nuclear physics \cite{Ring-Schuck}.
We expect that the mean-field approach for the Kondo effect in charm/bottom nuclei will give us a straightforward extension toward the impurity physics in nuclear theory.

The paper is organized as the followings.
In Section~2, we introduce the effective interaction for exchanging isospin between a $\bar{D}$, $B$ meson and a valence nucleon.
One of the purposes in this paper is to study the validity of the mean-field approximation for many-body problem in the Kondo effect, when the valence nucleons occupy the discrete energy-levels.
In Section~3, we investigate the approximate solution for the Kondo effect in the mean-field approximation.
We introduce the auxiliary fermion field for isospin of $\bar{D}$, $B$ meson, and apply the mean-field approximation in the extended Fock space.
We consider the quantum fluctuation by the random-phase approximation (RPA) and show that the approximate solution becomes closer to the exact one.
In Section~4, we discuss the competition between the Kondo effect and the correlations of valence nucleons.
We investigate the isospin breaking of the valence nucleons, and the nucleon pairings.
The final section is devoted for conclusion.

\section{Hamiltonian for Kondo effect with discrete energy-levels}

\subsection{Model setup}

We consider a $\bar{D}$, $B$ meson (isospin 1/2) as a heavy impurity particle in atomic nuclei.
In order to treat the isospin-exchange interaction between a $\bar{D}$, $B$ meson and a valence nucleon in a simple form as far as possible, we consider the Hamiltonian
\begin{eqnarray}
H = H_{0} + H_{\mathrm{K}},
\label{eq:H}
\end{eqnarray}
where $H_{0}$ is the kinetic term for the valence nucleon
\begin{eqnarray}
H_{0} =
\sum \epsilon_{k} {c_{k \sigma}}^{\!\dag} c_{k \sigma},
\label{eq:H0}
\end{eqnarray}
and $H_{\mathrm{K}}$ is the isospin-exchange (Kondo) interaction term
\begin{eqnarray}
H_{\mathrm{K}}
&=&
g  \sum  \left( {c_{k' \downarrow}}^{\!\dag} c_{k \uparrow} \, T_{+} + {c_{k' \uparrow}}^{\!\dag} c_{k \downarrow} \, T_{-} \right. \nonumber \\
&& \hspace{2.2em} \left. + ({c_{k' \uparrow}}^{\!\dag} c_{k \uparrow} \!-\! {c_{k' \downarrow}}^{\!\dag} c_{k \downarrow}) \, T_{3}  \right),
\label{eq:HK}
\end{eqnarray}
with the coupling constant $g$.
Here $c_{k\sigma}$ (${c_{k\sigma}}^{\!\dag}$) is the annihilation (creation) operator for the valence nucleon in the  $k$th single-state, where $k=1,\dots,N$ indicates the single-state of the valence nucleon\footnote{For example, $N$ is given by $N=2j+1$ for $j$-shell in nuclear shell model.}, and $\sigma= \uparrow$, $\downarrow$ is the up, down component of the isospin.
We define $T_{\pm}$ and $T_{3}$ as the raising/lowering operators and the $z$ component of the $\mathrm{SU}(2)$ isospin operator, and
\begin{eqnarray}
 T_{1} &=& \frac{1}{2} \left( T_{+} + T_{-} \right), \\
 T_{2} &=& \frac{1}{2i} \left( T_{+} - T_{-} \right).
\end{eqnarray}
$T_{1}$, $T_{2}$ and $T_{3}$ satisfy the commutation relation of the SU(2) algebra
\begin{eqnarray}
 \left[ T_{a}, T_{b} \right] = i \epsilon_{abc}T_{c},
\end{eqnarray}
with $a,b,c=1,2,3$.

We note that, in the Hamiltonian (\ref{eq:H}), 
there is a quantum fluctuation of the impurity isospin, because the direction of the impurity isospin is not fixed.
Therefore, the dynamics of the valence nucleon is always affected by the isospin fluctuation of the impurity, and hence it cannot be reduced to the one-body problem.
This is one of the interesting properties of the Kondo effect.
The purpose for us is to obtain the energy eigenvalues of the Hamiltonian (\ref{eq:H}) by considering the isospin fluctuation.

\subsection{Exact energy eigenvalues}
\label{sec:exact}

\subsubsection{Variational method for wave function}

\begin{table}[tdp]
{\small
\caption{Energy eigenvalues of the Hamiltonian (\ref{eq:H}) for $n=1$. The numbers in parentheses are the numbers of degeneracy factor.}
\begin{center}
\renewcommand{\arraystretch}{1.2}
\begin{tabular}{|c|c|c|}
\hline
 & \multicolumn{2}{|c|}{number of valence nucleon $n=1$} \\
\hline
$N$ & $I=0$ & $I=1$  \\
\hline
any $N$ & $\epsilon\!-\!\frac{3}{2}Ng$ (1), $\epsilon$ ($N\!-\!1$) & $\epsilon\!+\!\frac{1}{2}Ng$ (1), $\epsilon$ ($N\!-\!1$) \\
\hline
\end{tabular}
\end{center}
\label{table:exact_n=1}
}
\end{table}%

\begin{table}[tdp]
{\small
\caption{Energy eigenvalues of the Hamiltonian (\ref{eq:H}) for $n=2$. The notations are the same as Table~\ref{table:exact_n=1}.}
\begin{center}
\renewcommand{\arraystretch}{1.2}
\begin{tabular}{|c|c|c|}
\hline
 & \multicolumn{2}{|c|}{number of valence nucleon $n= 2$} \\
\hline
$N$ & $I=1/2$ & $I=3/2$  \\
\hline
$1$ & $2\epsilon$  (1)& ---  \\
\hline
$2$ & $2\epsilon\!-\!\frac{3}{2}Ng$ (1), $2\epsilon$ (2), $2\epsilon\!+\!\frac{1}{2}Ng$ (1) & \hspace{-3.4em} $2\epsilon\!+\!\frac{1}{2}Ng$ (1)  \\
\hline
$3$ & $2\epsilon\!-\!\frac{3}{2}Ng$ (2), $2\epsilon$ (5), $2\epsilon\!+\!\frac{1}{2}Ng$ (2) & $2\epsilon\!+\!\frac{1}{2}Ng$ (2), $2\epsilon$ (1) \\
\hline
$4$ & $2\epsilon\!-\!\frac{3}{2}Ng$ (3), $2\epsilon$ (10), $2\epsilon\!+\!\frac{1}{2}Ng$ (3) & $2\epsilon\!+\!\frac{1}{2}Ng$ (3), $2\epsilon$ (3)  \\
\hline
\end{tabular}
\end{center}
\label{table:exact_n=2}
}
\end{table}%

For simplicity, we consider the single-particle states with energy $\epsilon_{k}=\epsilon$ for the valence nucleons.
We use the representations $|\!\uparrow\,\rangle_{\mathrm{imp}}$ and $|\!\downarrow\,\rangle_{\mathrm{imp}}$ for the impurity states with isospin $\uparrow$ and $\downarrow$, respectively.
We also use the representation $|\psi^{(n)}_{I,I_{3}}\rangle$ for the total state, composed of an impurity and valence nucleons, with isospin $I$, its $z$ component $I_{3}$ and the number of valence nucleon $n$.
We consider as example the $n=1$, $2$, $3$ cases in the followings.

\vspace{0.5em}

\paragraph{The $n=1$ case.}
We consider isospin $I=0$, $1$.

For $I=0$, we assume the wave function 
\begin{eqnarray}
 |\psi^{(1)}_{0,0}\rangle
=
\sum \Gamma_{k} \left( {c_{k\uparrow}}^{\!\dag}|\!\downarrow\,\rangle_{\mathrm{imp}} - {c_{k\downarrow}}^{\!\dag}|\!\uparrow\,\rangle_{\mathrm{imp}} \right),
\end{eqnarray}
with unknown coefficients $\{\Gamma_{k}\}=\left\{ \Gamma_{1}, \dots, \Gamma_{N} \right\}$.
By using $H|\psi^{(1)}_{0,0}\rangle=E|\psi^{(1)}_{0,0}\rangle$, we obtain
\begin{eqnarray}
 \epsilon \, \Gamma_{k} - \frac{3}{2}g \sum \Gamma_{n} = E \, \Gamma_{k},
\end{eqnarray}
and hence the energy eigenvalues
\begin{eqnarray}
 E = \epsilon-\frac{3}{2}Ng\,\, (\mathrm{n.d.f.}=1), \hspace{0.5em} \epsilon\,\, (\mathrm{n.d.f.}=N-1).
\end{eqnarray}
The number in the parentheses are the number of degeneracy factor ($\mathrm{n.d.f.}$).

For $I=1$, considering $I_{3}=1$, we assume the wave function
\begin{eqnarray}
 |\psi^{(1)}_{1,1}\rangle
=
\sum \Gamma_{k} {c_{k\uparrow}}^{\!\dag}|\!\uparrow\,\rangle_{\mathrm{imp}} ,
\end{eqnarray}
with the unknown coefficients $\{\Gamma_{k}\}$.
By using $H|\psi^{(1)}_{1,1}\rangle=E|\psi^{(1)}_{1,1}\rangle$, we obtain the relation
\begin{eqnarray}
 \epsilon \, \Gamma_{k} + \frac{1}{2}g \sum \Gamma_{n} = E \, \Gamma_{k},
\end{eqnarray}
and hence the energy eigenvalues
\begin{eqnarray}
 E = \epsilon+\frac{1}{2}Ng\,\, (\mathrm{n.d.f.}=1), \hspace{0.5em} \epsilon\,\, (\mathrm{n.d.f.}=N-1).
\end{eqnarray}
We obtain the same values for $I_{3}=0$, $-1$.

See Table~\ref{table:exact_n=1} for summary.

\vspace{0.5em}

\paragraph{The $n=2$ case}
We consider $I=1/2$, $3/2$.

For $I=1/2$, considering $I_{3}=1/2$, we assume the wave function
\begin{widetext}
\begin{eqnarray}
|\psi^{(2)}_{1/2,1/2}\rangle 
 &=&
\sum \left\{ \Gamma_{mn}^{0} [{c_{m}}^{\!\dag} \!\otimes\! {c_{n}}^{\!\dag}]_{00} |\!\uparrow\,\rangle_{\mathrm{imp}}
+  \Gamma_{mn}^{1} \left(
        \sqrt{\frac{2}{3}} [{c_{m}}^{\!\dag} \!\otimes\! {c_{n}}^{\!\dag}]_{11} |\!\downarrow\,\rangle_{\mathrm{imp}}
      - \frac{1}{\sqrt{3}} [{c_{m}}^{\!\dag} \!\otimes\! {c_{n}}^{\!\dag}]_{10} |\!\uparrow\,\rangle_{\mathrm{imp}} \right) \right\}, 
\end{eqnarray}
\end{widetext}
with unknown coefficients $\Gamma_{mn}^{0}$, $\Gamma_{mn}^{1}$ having the properties $\Gamma_{mn}^{0}=\Gamma_{nm}^{0}$, $\Gamma_{mn}^{1}=-\Gamma_{nm}^{1}$.
Here we define
\begin{eqnarray}
[{c_{m}}^{\!\dag} \!\otimes\! {c_{n}}^{\!\dag}]_{00} = \frac{1}{\sqrt{2}} \left( {c_{m\uparrow}}^{\!\dag} {c_{n\downarrow}}^{\!\dag} - {c_{m\downarrow}}^{\!\dag} {c_{n\uparrow}}^{\!\dag} \right),
\end{eqnarray}
and 
\begin{eqnarray}
\hspace*{-1.5em}
[{c_{m}}^{\!\dag} \!\otimes\! {c_{n}}^{\!\dag}]_{1I_{3}} \!=\!
\left\{\!
\begin{array}{l}
 {c_{m\uparrow}}^{\!\dag} {c_{n\uparrow}}^{\!\dag} \hspace{0.2em} (I_{3}\!=\!1)  \\
 \frac{1}{\sqrt{2}} \left( {c_{m\uparrow}}^{\!\dag} {c_{n\downarrow}}^{\!\dag} \!+\! {c_{m\downarrow}}^{\!\dag} {c_{n\uparrow}}^{\!\dag} \right)  \hspace{0.2em} (I_{3}\!=\!0) \\
 {c_{m\downarrow}}^{\!\dag} {c_{n\downarrow}}^{\!\dag}  \hspace{0.2em} (I_{3}\!=\!-1)
\end{array}
\right.
\end{eqnarray}
for pairs of valence nucleons with isosinglet and isotriplet, respectively, for short notation.
By using $H|\psi^{(2)}_{1/2,1/2}\rangle=E|\psi^{(2)}_{1/2,1/2}\rangle$, we obtain the energy eigenvalues
\begin{eqnarray}
 E&=&\epsilon-\frac{3}{2}Ng \hspace{0.5em} (\mathrm{n.d.f.}\!=\!N\!-\!1), \hspace{0.5em}  \epsilon \hspace{0.5em} (\mathrm{n.d.f.}\!=\!N^2\!-\!2N\!+\!2),\nonumber \\
  && \epsilon+\frac{1}{2}Ng \hspace{0.5em} (\mathrm{n.d.f.}\!=\!N\!-\!1).
\end{eqnarray}
We obtain the same values for $I_{3}=-1/2$.

For $I=3/2$, considering $I_{3}=3/2$, we assume the wave function
\begin{eqnarray}
 | \psi^{(2)}_{3/2,3/2} \rangle 
 =
 \sum \Gamma_{mn}^{1} [{c_{m}}^{\!\dag} \!\otimes\! {c_{n}}^{\!\dag}]_{11} |\!\uparrow\,\rangle_{\mathrm{imp}},
\end{eqnarray}
with unknown coefficients $\Gamma_{mn}^{1}$ having the property $\Gamma_{mn}^{1}=-\Gamma_{nm}^{1}$.
By using $H|\psi^{(2)}_{3/2,3/2}\rangle=E|\psi^{(2)}_{3/2,3/2}\rangle$, we obtain, through the relation
\begin{eqnarray}
2\epsilon \, \Gamma_{mn}^{1} - \frac{1}{2} g\sum_{1 \le l \le N} \left( \Gamma_{lm}^{1} - \Gamma_{ln}^{1} \right) = E\, \Gamma_{mn}^{1},
\end{eqnarray}
with $m<n$, the energy eigenvalues
\begin{eqnarray}
 E&=&\epsilon \hspace{0.5em} \left( \mathrm{n.d.f.}=\frac{1}{2}(N-2)(N-1) \right), \nonumber \\
  && \epsilon+\frac{1}{2}Ng \hspace{0.5em} (\mathrm{n.d.f.}=N-1).
\end{eqnarray}
We obtain the same values for $I_{3}=1/2$, $0$, $-1/2$, $-3/2$.

See Table~\ref{table:exact_n=2} for summary.

\begin{table*}[tdp]
{\small
\caption{Energy eigenvalues of the Hamiltonian (\ref{eq:H}) for $n=3$. The notations are the same as Table~\ref{table:exact_n=1}.}
\begin{center}
\renewcommand{\arraystretch}{1.2}
\begin{tabular}{|c|c|c|c|}
\hline
 & \multicolumn{3}{|c|}{number of valence nucleon $n= 3$} \\
\hline
$N$ & $I=0$ & $I=1$ & $I=2$ \\
\hline
$1$ & --- & --- & --- \\
\hline
$2$ &  \hspace{-6em} $3\epsilon\!-\!\frac{3}{2}Ng$ (1), $3\epsilon$ (1) & \hspace{5.3em} $3\epsilon$ (1), $3\epsilon\!+\!\frac{1}{2}Ng$ (1) & --- \\
\hline
$3$ & $3\epsilon\!-\!\frac{3}{2}Ng$ (3), $3\epsilon$ (4), $3\epsilon\!+\!\frac{1}{2}Ng$ (1) & $3\epsilon\!-\!\frac{3}{2}Ng$ (1), $3\epsilon$ (3), $3\epsilon\!+\!\frac{1}{2}Ng$ (4) &  \hspace{2em} $3\epsilon+\frac{1}{2}Ng$ (1) \\
\hline
$4$ & $3\epsilon\!-\!\frac{3}{2}Ng$ (6), $3\epsilon$ (11), $3\epsilon\!+\!\frac{1}{2}Ng$ (3) & $3\epsilon\!-\!\frac{3}{2}Ng$ (3), $3\epsilon$ (12), $3\epsilon\!+\!\frac{1}{2}Ng$ (9) & $3\epsilon$ (1), $3\epsilon\!+\!\frac{1}{2}Ng$ (3) \\
\hline
\end{tabular}
\end{center}
\label{table:exact_n=3}
}
\end{table*}%

\vspace{0.5em}

\paragraph{The $n=3$ case}
We consider $I=0$, $1$, $2$.

For $I=0$, we assume the wave function
\begin{widetext}
\begin{eqnarray}
\hspace*{-2em}
 | \psi^{(3)}_{0,0} \rangle 
 &\!=\!&
\!\!\!\! \sum_{0\le l<m \le N} \!\!\!\! \Gamma_{lm} {c_{l\uparrow}}^{\!\dag} {c_{l\downarrow}}^{\!\dag} \frac{1}{\sqrt{2}} \left( {c_{m\uparrow}}^{\!\dag} |\!\downarrow\,\rangle_{\mathrm{imp}} \!-\! {c_{m\downarrow}}^{\!\dag} |\!\uparrow\,\rangle_{\mathrm{imp}} \right) 
+
\!\!\!\! \sum_{0\le l<m \le N} \!\!\!\! \Gamma'_{lm} {c_{m\uparrow}}^{\!\dag} {c_{m\downarrow}}^{\!\dag} \frac{1}{\sqrt{2}} \left( {c_{l\uparrow}}^{\!\dag} |\!\downarrow\,\rangle_{\mathrm{imp}} \!-\! {c_{l\downarrow}}^{\!\dag} |\!\uparrow\,\rangle_{\mathrm{imp}} \right) \nonumber \\
&&
+
\!\!\!\!\!\!\!\! \!\! \sum_{0\le l<m<n \le N} \!\!\!\!\!\!\!\! \!\! \Gamma_{lmn}^{00} \frac{1}{2} \left( {c_{l\uparrow}}^{\!\dag} {c_{m\downarrow}}^{\!\dag} \!-\! {c_{l\downarrow}}^{\!\dag} {c_{m\uparrow}}^{\!\dag} \right)  \left( {c_{n\uparrow}}^{\!\dag} |\!\downarrow\,\rangle_{\mathrm{imp}} \!-\! {c_{n\downarrow}}^{\!\dag} |\!\uparrow\,\rangle_{\mathrm{imp}} \right) \nonumber \\
&&+
\!\!\!\!\!\!\!\! \!\! \sum_{0\le l<m<n \le N} \!\!\!\!\!\!\!\!  \!\!  \Gamma_{lmn}^{11} \frac{1}{\sqrt{3}} \!\! \left\{ {c_{l\uparrow}}^{\!\dag} {c_{m\uparrow}}^{\!\dag} {c_{n\downarrow}}^{\!\dag} |\!\downarrow\,\rangle_{\mathrm{imp}} 
\!-\! \frac{1}{2} \left( {c_{l\uparrow}}^{\!\dag} {c_{m\downarrow}}^{\!\dag} \!+\! {c_{l\downarrow}}^{\!\dag} {c_{m\uparrow}}^{\!\dag} \right) 
 \!\! \left( {c_{n\uparrow}}^{\!\dag} |\!\downarrow\,\rangle_{\mathrm{imp}} \!+\! {c_{n\downarrow}}^{\!\dag} |\!\uparrow\,\rangle_{\mathrm{imp}} \right)
\!+\!{c_{l\downarrow}}^{\!\dag} {c_{m\downarrow}}^{\!\dag} {c_{n\uparrow}}^{\!\dag} |\!\uparrow\,\rangle_{\mathrm{imp}}
 \right\}, 
\end{eqnarray}
\end{widetext}
with unknown coefficients $\Gamma_{lm}$, $\Gamma'_{lm}$, $\Gamma_{lmn}^{00}$ and $\Gamma_{lmn}^{11}$ with $ l<m<n$.
By using $H|\psi^{(3)}_{0,0}\rangle=E|\psi^{(3)}_{0,0}\rangle$, we obtain the energy eigenvalues as shown in Table~\ref{table:exact_n=3}.

For $I=1$, considering $I_{3}=1$, we assume the wave function
\begin{widetext}
\begin{eqnarray}
 | \psi^{(3)}_{1,1} \rangle 
 &=&
\!\!\!\! \sum_{0\le l<m \le N} \!\!\!\! \Gamma_{lm} {c_{l\uparrow}}^{\!\dag} {c_{l\downarrow}}^{\!\dag} {c_{m\uparrow}}^{\!\dag} |\!\uparrow\,\rangle_{\mathrm{imp}} 
+
\!\!\!\! \sum_{0\le l<m \le N} \!\!\!\! \Gamma'_{lm} {c_{m\uparrow}}^{\!\dag} {c_{m\downarrow}}^{\!\dag}  {c_{l\uparrow}}^{\!\dag} |\!\uparrow\,\rangle_{\mathrm{imp}} \nonumber \\
&&+
\!\!\!\!\!\!\!\! \sum_{0\le l<m<n \le N} \!\!\!\!\!\!\!\!\Gamma_{lmn}^{01} \frac{1}{\sqrt{2}} \left( {c_{l\uparrow}}^{\!\dag} {c_{m\downarrow}}^{\!\dag} \!-\! {c_{l\downarrow}}^{\!\dag} {c_{m\uparrow}}^{\!\dag} \right)
 {c_{n\uparrow}}^{\!\dag}  |\!\uparrow\,\rangle_{\mathrm{imp}} 
+
\!\!\!\!\!\!\!\! \sum_{0\le l<m<n \le N} \!\!\!\!\!\!\!\! \Gamma_{lmn}^{10} {c_{l\uparrow}}^{\!\dag} {c_{m\uparrow}}^{\!\dag}
 \frac{1}{\sqrt{2}} \left( {c_{n\uparrow}}^{\!\dag} |\!\downarrow\,\rangle_{\mathrm{imp}} \!-\! {c_{n\downarrow}}^{\!\dag} |\!\uparrow\,\rangle_{\mathrm{imp}} \right) \nonumber \\
&&
+
\!\!\!\!\!\!\!\! \sum_{0\le l<m<n \le N} \!\!\!\!\!\!\!\! \Gamma_{lmn}^{11} 
\left\{ \frac{1}{\sqrt{2}} {c_{l\uparrow}}^{\!\dag} {c_{m\uparrow}}^{\!\dag} \frac{1}{\sqrt{2}} \left( {c_{n\uparrow}}^{\!\dag} |\!\downarrow\,\rangle_{\mathrm{imp}} \!+\! {c_{n\downarrow}}^{\!\dag} |\!\uparrow\,\rangle_{\mathrm{imp}} \right) 
\!-\! \frac{1}{\sqrt{2}} \frac{1}{\sqrt{2}} \left( {c_{l\uparrow}}^{\!\dag} {c_{m\downarrow}}^{\!\dag} \!+\! {c_{l\downarrow}}^{\!\dag} {c_{m\uparrow}}^{\!\dag} \right) {c_{n\uparrow}}^{\!\dag} |\!\uparrow\,\rangle_{\mathrm{imp}}
\right\},
\end{eqnarray}
\end{widetext}
with unknown coefficients $\Gamma_{lm}$, $\Gamma'_{lm}$, $\Gamma_{lmn}^{10}$ and $\Gamma_{lmn}^{11}$ with $l<m<n$.
By using $H|\psi^{(3)}_{1,1}\rangle=E|\psi^{(3)}_{1,1}\rangle$, we obtain the energy eigenvalues as shown in Table~\ref{table:exact_n=3}.
We obtain the same values for $I_{3}=0$, $-1$.

For $I=2$, considering $I_{3}=2$, we assume the wave function
\begin{eqnarray}
 | \psi^{(3)}_{2,2} \rangle 
 =
 \sum_{0\le l<m<n \le N} \Gamma_{lmn}^{11} {c_{l\uparrow}}^{\!\dag} {c_{m\uparrow}}^{\!\dag} {c_{n\uparrow}}^{\!\dag} |\!\uparrow\,\rangle_{\mathrm{imp}},
\end{eqnarray}
with unknown coefficients $\Gamma_{lmn}^{11}$ with $l<m<n$.
By using $H|\psi^{(3)}_{2,2}\rangle=E|\psi^{(3)}_{2,2}\rangle$, we obtain the energy eigenvalues as shown in Table~\ref{table:exact_n=3}.
We obtain the same values for $I_{3}=1$, $0$, $-1$, $-2$.

\subsubsection{Method by pseudo-isospin SU(2) algebra}

For the single-particle states with $\epsilon_{k}=\epsilon$, we obtain the energy eigenvalues specially by the simple method.
For this purpose, we define the operator 
\begin{eqnarray}
 C_{N \sigma} = \frac{1}{\sqrt{N}} \sum_{k=1}^{N} c_{k\sigma},
\end{eqnarray}
as a coherent sum of $c_{k\sigma}$.
This satisfies the commutation relation for fermions
\begin{eqnarray}
 \left\{ C_{N\sigma}, {C_{N\sigma'}}^{\!\dag} \right\} = \delta_{\sigma\sigma'}.
\end{eqnarray}
Defining the raising/lowering operators and the $z$ component
\begin{eqnarray}
 T^{\,\mathrm{c}}_{+} &=& {C_{N\uparrow}}^{\!\dag} C_{N\downarrow}, \\
 T^{\,\mathrm{c}}_{-} &=& {C_{N\downarrow}}^{\!\dag} C_{N\uparrow}, \\
 T^{\,\mathrm{c}}_{3} &=& \frac{1}{2} \left( {C_{N\uparrow}}^{\!\dag} C_{N\uparrow} - {C_{N\downarrow}}^{\!\dag} C_{N\downarrow} \right),
\end{eqnarray}
and $x$, $y$ components
\begin{eqnarray}
 T^{\mathrm{c}}_{1} &=& \frac{1}{2} \left( T^{\mathrm{c}}_{+} + T^{\mathrm{c}}_{-} \right), \\
 T^{\mathrm{c}}_{2} &=& \frac{1}{2i} \left( T^{\mathrm{c}}_{+} - T^{\mathrm{c}}_{-} \right),
\end{eqnarray}
we introduce the operator
\begin{eqnarray}
 \vec{T}^{\,\mathrm{c}} = \left( T^{\,\mathrm{c}}_{1},T^{\,\mathrm{c}}_{2},T^{\,\mathrm{c}}_{3} \right).
\end{eqnarray}
Those operators satisfy the SU(2) algebra; $ \left[ T^{\mathrm{c}}_{a}, T^{\mathrm{c}}_{b} \right] = i \epsilon_{abc} T^{\mathrm{c}}_{c} $ with $a,b,c=1,2,3$.
Hence we call $\vec{T}^{\,\mathrm{c}}$ the pseudo-isospin.
We distinguish this from the conventional isospin operator for each valence nucleon, because $\vec{T}^{\,\mathrm{c}}$ (or $C_{N\sigma}$) gives the coherent state of $k=1,\dots,N$ single-particle states.
We emphasize that the pseudo-isospin can be defined only when the single-particle states have the same energy ($\epsilon_{k}=\epsilon$)\footnote{We may note the SU(2) algebra holds for more general case,
\begin{eqnarray}
 T^{\,\mathrm{c}}_{+} &=& \frac{1}{N} \sum {c_{k'\uparrow}}^{\!\dag} {\cal S}_{k'k} c_{k\downarrow}, \\
 T^{\,\mathrm{c}}_{-} &=& \frac{1}{N} \sum {c_{k'\downarrow}}^{\!\dag} {\cal S}_{k'k} c_{k\uparrow}, \\
 T^{\,\mathrm{c}}_{3} &=& \frac{1}{2N} \sum \left( {c_{k'\uparrow}}^{\!\dag} {\cal S}_{k'k} c_{k\uparrow} - {c_{k'\downarrow}}^{\!\dag} {\cal S}_{k'k} c_{k\downarrow} \right),
\end{eqnarray}
with symmetric ${\cal S}_{k'k}$ (${\cal S}_{k'k}={\cal S}_{kk'}$).}.

By using the identity
\begin{eqnarray}
 T^{\,\mathrm{c}}_{+} \, T_{-} + T^{\,\mathrm{c}}_{-} \, T_{+} + 2T^{\,\mathrm{c}}_{3} \,  T_{3} = 2 \vec{T}^{\,\mathrm{c}} \!\cdot\! \vec{T},
\end{eqnarray}
we find that the interaction term (\ref{eq:HK}) 
 can be expressed as
\begin{eqnarray}
 H_{\mathrm{K}} = 2Ng \vec{T}^{\,\mathrm{c}} \!\cdot\! \vec{T}.
\end{eqnarray}
According to the compound isospin $|\vec{T}^{\,\mathrm{c}}+\vec{T}|=0$, $1$ (i.e. $\vec{T}^{\,\mathrm{c}} \cdot \vec{T}=-3/4$, $1/4$, respectively), we obtain
\begin{eqnarray}
 H_{\mathrm{K}} = 
\left\{
\begin{array}{ccc}
  -\frac{3}{2}Ng \hspace{1em} (|\vec{T}^{\,\mathrm{c}}+\vec{T}|=0) \\
 \frac{1}{2} Ng \hspace{1em} (|\vec{T}^{\,\mathrm{c}}+\vec{T}|=1) 
\end{array}
\right..
\end{eqnarray}
The original Hamiltonian (\ref{eq:H}) can be given as
\begin{eqnarray}
 H
 =
 \sum \epsilon \, {C_{k\sigma}}^{\!\dag} C_{k\sigma} + 2Ng \, \vec{T}^{\,\mathrm{c}} \!\cdot\! \vec{T},
\label{eq:H2}
\end{eqnarray}
by the fermion operator $C_{N\sigma}$ and the $N-1$ orthogonal  fermion operators $C_{k\sigma}$ ($k=1$, $\dots$, $N-1$).
$C_{k\sigma}$ with $k=1,\dots,N-1$ are linear combinations of $c_{k \sigma}$ with $k=1,\dots,N$, all of which are commutative with $C_{N\sigma}$, and satisfy $\{ {C_{k\sigma}}^{\!\dag}, C_{k'\sigma'} \} = \delta_{kk'}\delta_{\sigma\sigma'}$ ($k,k'=1,\dots,N-1$).
From Eq.~(\ref{eq:H2}), we can indeed confirm the results in Tables~\ref{table:exact_n=1}, \ref{table:exact_n=2}, \ref{table:exact_n=3}.

\section{Mean-field approximation in Kondo effect}

In the previous section, we obtained the energy eigenvalues by considering the simple case with $\epsilon_{k}=\epsilon$ for the Hamiltonian (\ref{eq:H}).
In general cases, however, we need to perform diagonalization of large matrices with paying a cost to the numerical computation.
Moreover, such direct analysis may not be useful for intuitive understanding of the result.
In this section, introducing the mean-field approach based on Ref.~\cite{Takano:1966,Yoshimori:1970,Eto:2001}, we discuss how this approximation brings an easy way to obtain the ground state of the Hamiltonian (\ref{eq:H}), and investigate the validity of the mean-field approach by comparing the results with the exact ones.
We also consider the quantum fluctuation in RPA beyond the mean-field approximation.

We note that the mean-field approach was applied to the cases with continuous number-density of valence fermions  in infinitely large system in condensed matter physics \cite{Takano:1966,Yoshimori:1970,Eto:2001}.
As emphasized in Introduction, the purpose in the present discussion is to investigate the Kondo effect in finite systems with discrete energy-levels of valence nucleons in charm/bottom nuclei.
For this purpose, we apply the mean-filed approach to the finite-size system with discrete energy-levels.
As analogy, we remember that the BCS theory, which is successful to describe the superconducting state with continuous number-density in infinitely large system, can be applied to pairings of valence nucleons in finite nuclei \cite{Ring-Schuck}.

\subsection{Introducing auxiliary fermion fields}

In order to describe the isospin of the impurity, we introduce the auxiliary fermion field $f_{\sigma}$ ($\sigma=\uparrow$, $\downarrow$) \cite{Takano:1966,Yoshimori:1970,Eto:2001}.
They satisfy the fermion commutation relation
$\{ f_{\sigma}, {f_{\sigma'}}^{\!\dag} \} = \delta_{\sigma \sigma'}$
and
$\left\{ f_{\sigma}, f_{\sigma'} \right\} = 0$.
We rewrite the isospin operators $T_{+}$, $T_{-}$, $T_{3}$ of the impurity by using $f_{\sigma}$ as
\begin{eqnarray}
T_{+} &=& {f_{\uparrow}}^{\!\dag} f_{\downarrow}, \label{eq:T_plus} \\
T_{-} &=& {f_{\downarrow}}^{\!\dag} f_{\uparrow}, \label{eq:T_minus} \\
T_{3} &=& \frac{1}{2} ( {f_{\uparrow}}^{\!\dag} f_{\uparrow} - {f_{\downarrow}}^{\!\dag} f_{\downarrow} ). \label{eq:T_z}
\end{eqnarray}
Because the number of the impurity should be always equal to one, we need to impose the constraint condition \cite{Takano:1966,Yoshimori:1970,Eto:2001}
\begin{eqnarray}
\sum {f_{\sigma}}^{\!\dag} f_{\sigma}=1.
\label{eq:constraint}
\end{eqnarray}
The Fock space satisfying this condition is the physical Fock space which should be obtained.
The Fock space with the other impurity numbers, $\sum {f_{\sigma}}^{\!\dag} f_{\sigma}=0$, $2$, which is indeed unphysical, needs to be excluded.
In the mean-field approximation, however, it will turn out that an extension of the Fock space to the multiple impurity-numbers is useful to analyze the ground state of the Hamiltonian (\ref{eq:H}).
In the followings, we consider separately the two cases of $g>0$ and $g<0$ in the Hamiltonian (\ref{eq:H}).

We note that the above decomposition of the operators $T_{+}$, $T_{-}$, $T_{3}$ can be given by boson fields instead of the fermion fields.
In the boson case, however, we need to consider superposed fields of the bosons and the valence nucleons, fermions, in the mean-field approximation, which may lead to some difficulty.
Moreover, the Fock space for the boson fields has to be extended to infinite number of bosons in contrast to the fermion case, where fermion numbers are limited to two at most.
Therefore, we consider that the fermion fields are more convenient than the boson fields in the present analysis.

\subsection{Isosinglet condensate ($g>0$)}
\label{sec:isosinglet_condensate}

We consider the $g>0$ case.
First we discuss the mean-field approximation, and second we investigate the fluctuation by using RPA. 

\subsubsection{Mean-field approximation}

We rewrite the Hamiltonian (\ref{eq:H}) as
\begin{eqnarray}
H
&=& \sum \epsilon_{k} {c_{k\sigma}}^{\!\dag} c_{k\sigma} \nonumber \\
&&+ g \left( \sum {f_{\sigma}}^{\!\dag} f_{\sigma'} {c_{k'\sigma'}}^{\!\dag} c_{k\sigma} - \frac{1}{2}  \sum {c_{k' \sigma}}^{\!\dag} c_{k \sigma} \right) \nonumber \\
&&+ \lambda \left( \sum {f_{\sigma}}^{\!\dag} f_{\sigma} -1 \right),
\label{eq:H_0_MF_1}
\end{eqnarray}
by using the relations (\ref{eq:T_plus})-(\ref{eq:T_z}) 
 and the identity
\begin{eqnarray}
&& {c_{k' \downarrow}}^{\!\dag} c_{k \uparrow} \, T_{+} + {c_{k' \uparrow}}^{\!\dag} c_{k \downarrow} \, T_{-} + ({c_{k' \uparrow}}^{\!\dag} c_{k \uparrow}-{c_{k' \downarrow}}^{\!\dag} c_{k \downarrow}) \, T_{3} \nonumber \\
&=&
\sum {f_{\sigma}}^{\!\dag} f_{\sigma'} {c_{k'\sigma'}}^{\!\dag} c_{k\sigma} - \frac{1}{2}  \sum {c_{k' \sigma}}^{\!\dag} c_{k \sigma},
\label{eq:identity}
\end{eqnarray}
where the constraint condition (\ref{eq:constraint}) is used \footnote{The second term in the right-hand side in Eq.~(\ref{eq:identity}) does not include the flipping of the isospin of the valence nucleon, and hence could be neglected for the Kondo effect \cite{Eto:2001}. However, we keep this term throughout the analysis, because the present discussion is devoted to comparison of the result in the mean-field approximation with the exact solution.}.
In the last term in the right-hand side in Eq.~(\ref{eq:H_0_MF_1}), we consider the constraint condition (\ref{eq:constraint}) by introducing the Lagrange multiplier constant $\lambda$.
Now we apply the mean-field approximation. We introduce the mean-field $\langle {f_{\sigma}}^{\!\dag} c_{k \sigma} \rangle$ as an expectation value of ${f_{\sigma}}^{\!\dag} c_{k \sigma}$, sandwiched by the ground state,
and define the isosinglet ``gap" function
\cite{Takano:1966,Yoshimori:1970,Eto:2001} 
\begin{eqnarray}
\Delta = -g \sum \langle {f_{\sigma}}^{\!\dag} c_{k \sigma} \rangle.
\label{eq:gap_0}
\end{eqnarray}
Using the relation
\begin{eqnarray}
&& g \sum {f_{\sigma}}^{\!\dag} f_{\sigma'} {c_{k'\sigma'}}^{\!\dag} c_{k\sigma} \nonumber \\
&=& g \sum {f_{\sigma}}^{\!\dag} f_{\sigma'} \left( -c_{k\sigma} {c_{k'\sigma'}}^{\!\dag} + \delta_{kk'} \delta_{\sigma \sigma'} \right) \nonumber \\
&=& -g \sum {f_{\sigma}}^{\!\dag} c_{k \sigma} \, {c_{k' \sigma'}}^{\!\dag} f_{\sigma'} + Ng \sum {f_{\sigma}}^{\!\dag} f_{\sigma} \nonumber \\
&=& -g \sum \left( {f_{\sigma}}^{\!\dag} c_{k \sigma} - \langle {f_{\sigma}}^{\!\dag} c_{k \sigma} \rangle + \langle {f_{\sigma}}^{\!\dag} c_{k \sigma} \rangle \right) \nonumber \\
&& \times \left( {c_{k' \sigma'}}^{\!\dag} f_{\sigma'} - \langle {c_{k' \sigma'}}^{\!\dag} f_{\sigma'} \rangle + \langle {c_{k' \sigma'}}^{\!\dag} f_{\sigma'} \rangle \right) + Ng \nonumber \\
&=&
-g \sum \left( {f_{\sigma}}^{\!\dag} c_{k \sigma} - \langle {f_{\sigma}}^{\!\dag} c_{k \sigma} \rangle \right) \left( {c_{k' \sigma'}}^{\!\dag} f_{\sigma'} - \langle {c_{k' \sigma'}}^{\!\dag} f_{\sigma'} \rangle \right) \nonumber \\
&&-
g \sum \left( \langle {f_{\sigma}}^{\!\dag} c_{k \sigma} \rangle {c_{k' \sigma'}}^{\!\dag} f_{\sigma'} + \langle {c_{k' \sigma'}}^{\!\dag} f_{\sigma'} \rangle {f_{\sigma}}^{\!\dag} c_{k \sigma} \right) \nonumber \\
&&+ g \sum \langle {f_{\sigma}}^{\!\dag} c_{k \sigma} \rangle \langle {c_{k' \sigma'}}^{\!\dag} f_{\sigma'} \rangle + Ng,
\end{eqnarray}
where the constraint condition (\ref{eq:constraint}) is used again,
we separate the Hamiltonian (\ref{eq:H}) into the mean-field part $H_{\mathrm{MF}}$ and the fluctuation part $H_{\mathrm{fluc}}$ as
\begin{eqnarray}
H = H_{\mathrm{MF}} + H_{\mathrm{fluc}},
\end{eqnarray}
with
\begin{eqnarray}
H_{\mathrm{MF}}
&=&
\sum \epsilon_{k} {c_{k\sigma}}^{\!\dag}c_{k\sigma} + \sum \left( \Delta^{\ast} {f_{\sigma}}^{\!\dag} c_{k\sigma} + \Delta {c_{k \sigma}}^{\!\dag} f_{\sigma} \right) \nonumber \\
&&+ \lambda \sum {f_{\sigma}}^{\!\dag} f_{\sigma} + \frac{|\Delta|^{2}}{g} - \lambda,
\end{eqnarray}
and
\begin{eqnarray}
H_{\mathrm{fluc}}
\!&=&\!
-g \sum \left( {f_{\sigma}}^{\!\dag} c_{k \sigma} \!-\! \langle {f_{\sigma}}^{\!\dag} c_{k \sigma} \rangle \right) \! \left( {c_{k'\sigma'}}^{\!\dag} f_{\sigma'} \!-\! \langle {c_{k'\sigma'}}^{\!\dag} f_{\sigma'} \rangle \right) \nonumber \\
&&- \frac{1}{2} g \sum {c_{k'\sigma}}^{\!\dag} c_{k \sigma} + Ng.
 \label{eq:H_0_fluc}
\end{eqnarray}

In the man-field approximation, we consider only the mean-field part $H_{\mathrm{MF}}$ and neglect the fluctuation part $H_{\mathrm{fluc}}$ \cite{Takano:1966,Yoshimori:1970,Eto:2001}.
We diagonalize $H_{\mathrm{MF}}$ and introduce the Slater determinant by single-particle states.
Then, we perform the variation for the expectation value $\langle H_{\mathrm{MF}} \rangle$ with respect to $\lambda$ and $\Delta$ as
\begin{eqnarray}
\frac{\partial}{\partial \lambda} \langle H_{\mathrm{MF}} \rangle &=& 0, \\
\frac{\partial}{\partial \Delta} \langle H_{\mathrm{MF}} \rangle &=& 0,
\end{eqnarray}
and finally obtain $\lambda$ and $\Delta$.
The ground-state energy is given by substituting the $\lambda$ and $\Delta$ into $\langle H_{\mathrm{MF}} \rangle$.

In the following, to demonstrate the mean-field calculation explicitly, we consider the simple case of $\epsilon_{k}=\epsilon$ for all $k=1,\dots,N$, because the diagonalization of $H_{\mathrm{MF}}$ can be analytically performed.
Such simplification does not change the essence of the discussion.
With the basis $\{c_{k \sigma},f_{\sigma}\}$ ($k=1$, $\dots$, $N$, $\sigma = \uparrow$, $\downarrow$), we give the mean-field Hamiltonian $H_{\mathrm{MF}}$ in terms of the $2(N+1) \times 2(N+1)$ matrix ${\cal H}_{cf}$,
\begin{eqnarray}
{\cal H}_{cf}
=
\left(
\begin{array}{cccc|cccccc}
 \epsilon & 0 & \cdots  & \Delta^{\ast} & 0 & 0 &  \cdots & 0 \\
 0 & \epsilon & \cdots  & \Delta^{\ast} & 0 & 0 & \cdots & 0 \\
 \vdots & \vdots  & \ddots  & \vdots & \vdots & \vdots &  \ddots & \vdots \\
 \Delta & \Delta & \cdots & \lambda & 0 & 0 & \cdots & 0 \\  
 \hline
 0 & 0 &  \cdots & 0 & \epsilon & 0 & \cdots  & \Delta^{\ast} \\
 0 & 0 & \cdots & 0 & 0 & \epsilon & \cdots  & \Delta^{\ast} \\
 \vdots & \vdots &  \ddots & \vdots & \vdots & \vdots  & \ddots  & \vdots \\
 0 & 0 & \cdots & 0 &  \Delta & \Delta & \cdots & \lambda \\
\end{array}
\right),
 \label{eq:H0_matrix}
\end{eqnarray}
as
\begin{eqnarray}
H_{\mathrm{MF}} 
=
\psi^{\dag }{\cal H}_{cf}
\psi
+\frac{|\Delta|^{2}}{g} - \lambda,
\end{eqnarray}
with defining
\begin{eqnarray}
\psi
=
\left(
\begin{array}{c}
 c_{1 \uparrow} \\
 \vdots \\
 c_{N\uparrow} \\
 f_{\uparrow} \\  
 \hline
 c_{1 \downarrow} \\
 \vdots \\
 c_{N\downarrow} \\
 f_{\downarrow} \\
\end{array}
\right),
\label{eq:psi_field}
\end{eqnarray}
for short notation.
It is worth to note that $g>0$ should be maintained, because the stability of the ground state is guaranteed by the positivity of $|\Delta|^{2}/g$ in $H_{\mathrm{MF}}$.
Then, we diagonalize ${\cal H}_{cf}$ analytically as
\begin{widetext}
\begin{eqnarray}
{\cal H}_{cf}^{\mathrm{diag}}
&=&
\mathrm{diag} \left( \epsilon, \dots, \epsilon, \frac{1}{2} (\epsilon+\lambda-D), \frac{1}{2} (\epsilon+\lambda+D),
\epsilon, \dots, \epsilon, \frac{1}{2} (\epsilon+\lambda-D), \frac{1}{2} (\epsilon+\lambda+D) \right) \nonumber \\
&=&
\mathrm{diag} \left( E_{1},  \dots, E_{N-1}, E_{N}, E_{N+1}, E_{1}, \dots, E_{N-1}, E_{N}, E_{N+1} \right),
\label{eq:H_MF_0}
\end{eqnarray}
\end{widetext}
with
\begin{eqnarray}
 D=\sqrt{(\epsilon-\lambda)^{2}+4N|\Delta|^{2}}.
\end{eqnarray}
Introducing the new fields $\{d_{k \sigma}\}$ ($k=1$, $\dots$, $N$)
\begin{eqnarray}
d_{1\sigma} &=& \frac{1}{\sqrt{2}} \left( c_{1\sigma}-c_{2\sigma} \right), \\
d_{2\sigma} &=& \frac{1}{\sqrt{2}} \left( c_{1\sigma}-c_{3\sigma} \right),  \\
&\vdots& \nonumber \\
d_{N-1\sigma} &=& \frac{1}{\sqrt{2}} \left(c_{1\sigma}-c_{N\sigma}\right),\\
d_{N\sigma} &=& \frac{1}{\sqrt{2N}} \sqrt{1-\frac{\epsilon-\lambda}{D}} \left( c_{1\sigma}+\dots+c_{N\sigma} \right) \nonumber \\
&&
- \frac{1}{\sqrt{2}} \sqrt{1+\frac{\epsilon-\lambda}{D}} f_{\sigma}, \\
d_{N+1\sigma} &=& \frac{1}{\sqrt{2N}} \sqrt{1+\frac{\epsilon-\lambda}{D}} \left( c_{1\sigma}+\dots+c_{N\sigma} \right) \nonumber \\
&& + \frac{1}{\sqrt{2}} \sqrt{1-\frac{\epsilon-\lambda}{D}} f_{\sigma}, 
\end{eqnarray}
we represent the mean-field Hamiltonian $H_{\mathrm{MF}}$ by
\begin{eqnarray}
H_{\mathrm{MF}}
&=&
\phi^{\dag} {\cal H}_{cf}^{\mathrm{diag}}
\phi
+ \frac{|\Delta|^{2}}{g} - \lambda \nonumber \\
&=&
\sum E_{k} {d_{k \sigma}}^{\!\dag} d_{k \sigma} + \frac{|\Delta|^{2}}{g} - \lambda,
\label{eq:Ham_MF_0}
\end{eqnarray}
with defining
\begin{eqnarray}
\phi
=
\left(
\begin{array}{c}
 d_{1 \uparrow} \\
 \vdots \\
 d_{N\uparrow} \\
 d_{N+1\uparrow} \\  
 \hline
 d_{1 \downarrow} \\
 \vdots \\
 d_{N\downarrow} \\
 d_{N+1\downarrow} \\
\end{array}
\right).
\label{eq:phi_field}
\end{eqnarray}
 We remark that the isospin components  $\uparrow$ and  $\downarrow$ for the valence nucleons are separated in the matrix ${\cal H}_{cf}$,
  and the mixing part in the off-diagonal components is absorbed into the fluctuation part $H_{\mathrm{fluc}}$.
This separation indeed enables us to introduce the mean field for the valence nucleons.

Now let us consider the variation of $\langle H_{\mathrm{MF}} \rangle$ with respect to $\lambda$ and $\Delta$.
As a simple case, we consider the system with one valence nucleon.
The extension to $n$ valence nucleons is straightforward as discussed later.
In the present case, we have two degrees of freedom; an impurity and a valence nucleon.
To describe this system by the fields $d_{N\uparrow}$ and $d_{N\downarrow}$ having the minimum energy $E_{N}$, we consider the ground state
\begin{eqnarray}
| \psi_{0} \rangle = {d_{N \uparrow}}^{\!\dag} {d_{N \downarrow}}^{\!\dag} | 0 \rangle,
\end{eqnarray}
as the most stable state.
Performing the variation for
\begin{eqnarray}
E_{\mathrm{MF}}(\lambda,\Delta) &=& \langle \psi_{0} | H_{\mathrm{MF}} | \psi_{0} \rangle \nonumber \\
&=& 2E_{N} + \frac{|\Delta|^{2}}{g} - \lambda
\end{eqnarray}
with respect to $\lambda$ and $\Delta$,
\begin{eqnarray}
 \frac{\partial}{\partial \lambda} E_{\mathrm{MF}} =0, \hspace{1em}  \frac{\partial}{\partial \Delta} E_{\mathrm{MF}} =0,
\end{eqnarray}
we obtain the values of $\lambda$ and $\Delta$
\begin{eqnarray}
\lambda = \epsilon, \hspace{1em} \Delta = \sqrt{N}g.
\end{eqnarray}
The ground-state energy for the mean-field Hamiltonian $H_{\mathrm{MF}}$ is
\begin{eqnarray}
 E_{\mathrm{MF}}(\epsilon,\sqrt{N}g) = \epsilon - Ng.
\end{eqnarray}
Because we need to consider the energy shift $\langle \psi_{0} | H_{\mathrm{fluc}} | \psi_{0} \rangle=0$ by the fluctuation part $H_{\mathrm{fluc}}$, we finally obtain the ground-state energy for the original Hamiltonian (\ref{eq:H})
\begin{eqnarray}
  E_{\mathrm{MF}+\mathrm{shift}} 
&=& \langle \psi_{0} | H_{\mathrm{MF}} | \psi_{0} \rangle + \langle \psi_{0} | H_{\mathrm{fluc}} | \psi_{0} \rangle \nonumber \\
&=& E_{\mathrm{MF}}(\epsilon,\sqrt{N}g) + 0 \nonumber \\
&=&  \epsilon - Ng,
\label{eq:mean-field_energy_0}
\end{eqnarray}
in the mean-field approximation.
The binding energy $-Ng$ is different by about 33\% in contrast to the exact value $E_{\mathrm{exact}}=\epsilon - 3 Ng/2$ in Section~\ref{sec:exact}.
This difference originates from the limit of the mean-field approximation.
We expect that the correction by the fluctuation, which is not included in the mean-field approximation, enables us to get the value close to the exact one.
In the next subsection, we will discuss the energy correction by RPA. We furthermore discuss the result when the fluctuation is completely included in Appendix~\ref{sec:complete_fluctuation}.

We leave a comment on the obtained wave function $|\psi_{0}\rangle$.
Representing $|\psi_{0}\rangle$ by the original fields $\{c_{k\sigma},f_{\sigma}\}$, we find that $|\psi_{0}\rangle$ is a superposition of multiple number of impurities, i.e. $\sum {f_{\sigma}}^{\!\dag}f_{\sigma}=0$, $1$, $2$.
However, we should remind us that only one impurity is allowed to exist due to the condition (\ref{eq:constraint}).
In fact, we confirm this is satisfied as average by
\begin{eqnarray}
 \langle \psi_{0} | \sum {f_{\sigma}}^{\!\dag} f_{\sigma} | \psi_{0} \rangle = 1,
\end{eqnarray}
in the present mean-field approximation \cite{Takano:1966,Yoshimori:1970,Eto:2001}.
We also note that the ground state $|\psi_{0}\rangle$ is a state superposed coherently by many states of valence nucleon $k=1,\dots,N$.

We also leave a comment about the gap function (\ref{eq:gap_0}).
In the mean-field approximation, we introduced the new fields $\{d_{k\sigma}\}$ and considered the single-particle state for them.
In this basis, the gap function gives the strength of the binding energy in the system.
On the other hand, in the original fields $\{c_{k\sigma},f_{\sigma}\}$, the gap function gives the strength of the state mixing between the valence nucleon ($c_{k\sigma}$) and the impurity ($f_{\sigma}$) as seen in the matrix (\ref{eq:H0_matrix}) (see also Refs.~\cite{Takano:1966,Yoshimori:1970,Eto:2001}).
Although, the gap function gives the different physical meaning (the binding energy or the strength of the state mixing) according to the difference of the basis fields, they give essentially the same result.

\subsubsection{Fluctuation effect ---RPA---}

The mean-field approximation does not include the fluctuation effect.
In this subsection, we investigate the fluctuation effect based on RPA \cite{Ring-Schuck,Rowe:1968zza} (see also Refs.~\cite{Hagino:1999mu,Hagino:2000ba} for application to the Hartree-Fock states and the BCS states in atomic nuclei).
We rewrite the Hamiltonian (\ref{eq:H}) 
 in terms of $\{d_{k \sigma}\}$ instead of $\{c_{k\sigma},f_{\sigma}\}$ as
\begin{eqnarray}
H
&=&
\left(\epsilon - \frac{3}{4}Ng\right) \nonumber \\
&& \times \left\{ ( {a_{0 \uparrow}}^{\!\dag}a_{0 \uparrow} + {a_{0 \downarrow}}^{\!\dag}a_{0 \downarrow})+ ( {a_{1 \uparrow}}^{\!\dag}a_{1 \uparrow} + {a_{1 \downarrow}}^{\!\dag}a_{1 \downarrow}) \right\} \nonumber \\
&&+
\frac{1}{4} Ng \left \{ ({a_{0 \uparrow}}^{\!\dag} a_{1 \uparrow}+{a_{0 \downarrow}}^{\!\dag} a_{1 \downarrow})+({a_{1 \uparrow}}^{\!\dag} a_{0 \uparrow}+{a_{1 \downarrow}}^{\!\dag} a_{0 \downarrow}) \right \} \nonumber \\
&&+
\frac{1}{2}Ng \left( {a_{0\uparrow}}^{\!\dag}{a_{0\downarrow}}^{\!\dag} - {a_{1\uparrow}}^{\!\dag}{a_{1\downarrow}}^{\!\dag} \right)
                      \left( a_{0\uparrow}a_{0\downarrow} - a_{1\uparrow}a_{1\downarrow} \right) \nonumber \\
&&+
(-1)Ng \left( {a_{0\uparrow}}^{\!\dag}{a_{1\uparrow}}^{\!\dag} a_{0\uparrow}a_{1\uparrow} + {a_{0\downarrow}}^{\!\dag}{a_{1\downarrow}}^{\!\dag} a_{0\downarrow}a_{1\downarrow} \right) \nonumber \\
&&+
\left(-\frac{1}{2}\right) Ng \left( {a_{0\uparrow}}^{\!\dag}{a_{1\downarrow}}^{\!\dag} + {a_{0\downarrow}}^{\!\dag}{a_{1\uparrow}}^{\!\dag} \right)
                      \left( a_{0\uparrow}a_{1\downarrow} + a_{0\downarrow}a_{1\uparrow} \right) \nonumber \\
&&+
\left( -\epsilon + Ng \right) \nonumber \\
&&+
\sum_{k=1}^{N-1} E_{k} {d_{k \sigma}}^{\!\dag} d_{k \sigma},
\label{eq:H_RPA_0}       
\end{eqnarray}
where we define
$a_{0\sigma} = d_{N \sigma}$ and 
$a_{1\sigma} = d_{N+1 \sigma}$ 
for short notation.
Now we consider the RPA correlation energy by using the ground state $|\psi_{0}\rangle={a_{0 \uparrow}}^{\!\dag}{a_{0 \downarrow}}^{\!\dag} | 0 \rangle$ in the mean-field approximation.

First of all, we calculate energy eigenvalues of the RPA modes.
W consider the fluctuation near the ground state $|\psi_{0}\rangle={a_{0 \uparrow}}^{\!\dag}{a_{0 \downarrow}}^{\!\dag} | 0 \rangle$.
We solve the RPA equation
\begin{eqnarray}
\left(
\begin{array}{cc}
 A & B  \\
 -B^{\ast} & -A^{\ast}
\end{array}
\right)
\left(
\begin{array}{c}
 X  \\
 Y
\end{array}
\right)
=
\Omega_{\nu}
\left(
\begin{array}{c}
 X  \\
 Y
\end{array}
\right),
\end{eqnarray}
with
\begin{eqnarray}
\hspace*{-2em}
A_{\mu\nu\rho\sigma} &=&  \langle \psi_{0} | \left[ {a_{0\nu}}^{\!\dag} a_{1\mu}, \left[ H, {a_{1\rho}}^{\!\dag} a_{0\sigma} \right] \right]  | \psi_{0} \rangle, \nonumber \\
&=& \frac{1}{2} Ng \delta_{\mu\rho} \delta_{\nu\sigma} \nonumber \\
&&
 + Ng \left( \delta_{\mu\uparrow} \delta_{\nu\downarrow} \delta_{\rho\uparrow} \delta_{\sigma\downarrow} + \delta_{\mu\downarrow} \delta_{\nu\uparrow} \delta_{\rho\downarrow} \delta_{\sigma\uparrow} \right)  \nonumber \\
 && + \frac{1}{2} Ng \left( \delta_{\mu\uparrow} \delta_{\nu\uparrow} - \delta_{\mu\downarrow} \delta_{\nu\downarrow} \right) \left( \delta_{\rho\uparrow} \delta_{\sigma\uparrow} - \delta_{\rho\downarrow} \delta_{\sigma\downarrow} \right),
\end{eqnarray}
and
\begin{eqnarray}
\hspace*{-2em}
-B_{\mu\nu\rho\sigma} &=& \langle \psi_{0} | \left[ {a_{0\nu}}^{\!\dag} a_{1\mu}, \left[ H, {a_{0\sigma}}^{\!\dag} a_{1\rho} \right] \right]  | \psi_{0} \rangle \nonumber \\
&=&
\frac{1}{2} Ng \left( \delta_{\mu\uparrow}\delta_{\rho\downarrow} - \delta_{\mu\downarrow} \delta_{\rho\uparrow} \right) \left( \delta_{\nu\uparrow} \delta_{\sigma\downarrow} - \delta_{\nu\downarrow}\delta_{\sigma\uparrow} \right),
\end{eqnarray}
and obtain the RPA energy eigenvalues
\begin{eqnarray}
\left\{ \Omega_{\nu} \right\}
&=& \left\{ \Omega_{\pm1},\Omega_{\pm2},\Omega_{\pm3},\Omega_{0} \right\} \nonumber \\
&=& \left\{\pm \sqrt{2} Ng,\pm \sqrt{2} Ng,\pm \sqrt{2} Ng,0\right\}.
\end{eqnarray}
The zero-energy mode with $\Omega_{0}=0$ is due to the energy degeneracy of the first term in the Hamiltonian (\ref{eq:H_RPA_0}) for $| \psi_{0} \rangle = {a_{0\uparrow}}^{\!\dag}{a_{0\downarrow}}^{\!\dag} | 0 \rangle$ and $| \psi_{1} \rangle = {a_{1\uparrow}}^{\!\dag}{a_{1\downarrow}}^{\!\dag} | 0 \rangle$. 
This degeneracy is special in the mean-field approximation, and hence should be regarded as the spurious one.
Indeed, we will see such degeneracy will be resolved when higher order fluctuations are included in Appendix.~\ref{sec:complete_fluctuation}.

From the above result, we obtain the RPA correlation energy \cite{Ring-Schuck,Rowe:1968zza,Hagino:1999mu,Hagino:2000ba}
\begin{eqnarray}
\Delta E_{\mathrm{RPA}}
&=&
\frac{1}{2} \sum_{\nu>0} \Omega_{\nu} - \frac{1}{2} \mathrm{Tr}A \nonumber \\
&=&
\frac{1}{2} (3\sqrt{2}-5) Ng \nonumber \\
&\simeq& -0.378 Ng,
\end{eqnarray}
as the energy difference between the mean-field state and the fluctuating state.
Thus, the RPA correlation energy gives the correction to the ground-state energy in the mean-field approximation.
Therefore, the ground-state energy in the mean-field approximation and the RPA is 
\begin{eqnarray}
E_{\mathrm{MF}+\mathrm{shift}+\mathrm{RPA}}
&=&
E_{\mathrm{MF}+\mathrm{shift}}+\Delta E_{\mathrm{RPA}} \nonumber \\
&=&
\epsilon - \frac{1}{2} (7-3\sqrt{2}) Ng \nonumber \\
&\simeq&
\epsilon - 1.378 Ng.
\end{eqnarray}
This is the result for one valence nucleon.
For $n$ valence nucleons ($n \le 2N$), one nucleon participates in the binding as described above and the left $n-1$ valence nucleons does not (see Eq.~(\ref{eq:H_MF_0})).
Therefore, the energy becomes
\begin{eqnarray}
E_{\mathrm{MF}+\mathrm{shift}+\mathrm{RPA}}(n)
&\simeq&
n \, \epsilon - 1.378 Ng.
\end{eqnarray}
The binding energy $-1.378Ng$ is about 92\% of the exact solution $-3Ng/2$ in Section~\ref{sec:exact}.
Thus, by including the fluctuation in the RPA, we get the energy close to the exact one.
We expect that more closer value can be obtained when higher order fluctuations are taken into account.
In fact, we can diagonalize completely the Hamiltonian (\ref{eq:H_RPA_0}), due to its simplicity, and obtain the ground-state energy which is precisely the same as the exact one as presented in Appendix~\ref{sec:complete_fluctuation}.

\subsubsection{Correspondence between exact solution and mean-field+RPA solution}

Let us see the correspondence between the mean-field+RPA solution and the exact solution (Tables~\ref{table:exact_n=1}, \ref{table:exact_n=2}, \ref{table:exact_n=3}; $g>0$).
Concerning the ground state, we find that the former reproduces the latter within the approximation.

We consider the $n=1$ case.
For $N$ single-particle states of valence nucleon, we have one single-particle state which is coupled to impurity (coupling orbital) and $N-1$ single-particle states which are not coupled (non-coupling orbital).
In the ground state, one valence nucleon occupies the coupling orbital, and forms the isosinglet state as combined to the impurity isospin as the most stable state.
Therefore, the number of degeneracy factor is one.
This corresponds to the ground state of $I=0$, $1$ with energy $\epsilon-3Ng/2$ in Table~\ref{table:exact_n=1}.

We consider the $n=2$ case.
In this case, one of the two valence nucleons occupies the coupling orbital, and forms the isosinglet state combined with the impurity isospin.
The left valence nucleon occupies one of the $N-1$ non-coupling orbitals.
Because the fist valence nucleon forms the isosinglet state with the impurity, the addition of the second valence nucleon gives isodoublet state.
The number of degeneracy factor is $N-1$.
This is the same as the number of degeneracy factor in the $I=1/2$ ground state with energy $2\epsilon-3Ng/2$ in Table~\ref{table:exact_n=2}.

We consider the $n=3$ case.
In this case, one valence nucleon occupies the coupling orbital and forms the isosinglet state combined with the impurity isospin.
The other two valence nucleons occupy one or two of the $N-1$ non-coupling orbitals, and form the isosinglet or isotriplet state.
For the isosinglet state, the number of degeneracy factor is $N(N-1)/2$, because the second two valence nucleons can occupy the same single-particle states ($N-1$ patterns) or can occupy the different single-particle states ($(N-1)(N-2)/2$ patterns).
For the isotriplet state, the number of degeneracy factor is $(N-1)(N-2)/2$, because the two valence nucleons should occupy the different single-particle states ($(N-1)(N-2)/2$ patterns).
We confirm those numbers of degeneracy factor is consistent with those in the ground state of $I=0$, $1$ with energy $3\epsilon-3Ng/2$ in Table~\ref{table:exact_n=3}.

\subsection{Isotriplet condensate ($g<0$)}
\label{sec:isotriplet_condensate}

As mentioned previously, there is no stable isosinglet condensate for $g<0$.
In this case, we need to consider the isotriplet condensate.

\subsubsection{Mean-field approximation}

We rewrite the Hamiltonian (\ref{eq:H}) as
\begin{eqnarray}
H
&=&
\sum \epsilon_{k} {c_{k\sigma}}^{\!\dag} c_{k\sigma} \nonumber \\
&&
+g \sum {f_{\sigma}}^{\!\dag} (\sigma^{i})_{\sigma\rho} c_{k\rho} {c_{k'\sigma'}}^{\!\dag} (\sigma^{i})_{\sigma'\rho'} f_{\rho'} \nonumber \\
&&
+\frac{3}{2} g \sum {c_{k'\sigma}}^{\!\dag} c_{k\sigma} - 3Ng,
\end{eqnarray}
by using the identity
\begin{eqnarray}
&& \sum \left\{ {c_{k'\downarrow}}^{\!\dag} c_{k\uparrow} T_{+} \!+\! {c_{k'\uparrow}}^{\!\dag} c_{k\downarrow} T_{-} \!+\! \left( {c_{k'\uparrow}}^{\!\dag} c_{k\uparrow} \!-\! {c_{k'\downarrow}}^{\!\dag} c_{k\downarrow} \right) T_{3} \right\} \nonumber \\
&=&
\sum {f_{\sigma}}^{\!\dag} (\sigma^{i})_{\sigma\rho} c_{k\rho} {c_{k'\sigma'}}^{\!\dag} (\sigma^{i})_{\sigma'\rho'} f_{\rho'}
+\frac{3}{2} \sum {c_{k'\sigma}}^{\!\dag} c_{k\sigma} \nonumber \\
&&
- 3N, 
\end{eqnarray}
where the constraint condition (\ref{eq:constraint}) is used.
Defining the isotriplet ``gap" function
\begin{eqnarray}
 \Delta^{i} = g \sum \langle {f_{\alpha}}^{\!\dag} (\sigma^{i})_{\alpha\beta} c_{k\beta} \rangle,
 \label{eq:gap_1}
\end{eqnarray}
we separate the Hamiltonian (\ref{eq:H}) into the mean-field part $H'_{\mathrm{MF}}$ and the fluctuation part $H'_{\mathrm{fluc}}$
\begin{eqnarray}
 H = H'_{\mathrm{MF}} + H'_{\mathrm{fluc}},
\end{eqnarray}
with
\begin{eqnarray}
 H'_{\mathrm{MF}}
 &=&
 \sum \epsilon_{k} {c_{k\sigma}}^{\!\dag} c_{k\sigma} \nonumber \\
 &&+ \sum \left( \Delta^{i} {c_{k\sigma}}^{\!\dag} (\sigma^{i}_{\sigma\rho}) f_{\rho} + \Delta^{i\ast} {f_{\sigma}}^{\!\dag} (\sigma^{i}_{\sigma\rho}) c_{k\rho} \right) \nonumber \\
&&
 - \frac{1}{g} \sum |\Delta^{i}|^{2},
\end{eqnarray}
and
\begin{eqnarray}
 H'_{\mathrm{fluc}}
 &=&
g \sum {f_{\sigma}}^{\!\dag} (\sigma^{i})_{\sigma\rho} c_{k\rho} {c_{k'\sigma'}}^{\!\dag} (\sigma^{i})_{\sigma'\rho'} f_{\rho'} \nonumber \\
&&
- \sum \left( \Delta^{i} {c_{k\sigma}}^{\!\dag} (\sigma^{i}_{\sigma\rho}) f_{\rho} + \Delta^{i\ast} {f_{\sigma}}^{\!\dag} (\sigma^{i}_{\sigma\rho}) c_{k\rho} \right) \nonumber \\
&& + \frac{1}{g} \sum |\Delta^{i}|^{2}
+ \frac{3}{2} \sum {c_{k'\sigma}}^{\!\dag} c_{k\sigma} - 3Ng.
\end{eqnarray}
We note that $\Delta^{i}$ is given by the matrix form
\begin{eqnarray}
 \Delta^{i} (\sigma^{i})_{\alpha\beta}
 =
\left(
\begin{array}{cc}
 \Delta^{3} & \Delta^{1}-i\Delta^{2}  \\
 \Delta^{1}+i\Delta^{2} & -\Delta^{3}
\end{array}
\right)_{\alpha\beta}.
\end{eqnarray}

In the mean-field approximation, we consider only the mean-field part $H'_{\mathrm{MF}}$ and neglect the fluctuation part $H'_{\mathrm{fluc}}$, as performed in the isosinglet condensate in Section~\ref{sec:isosinglet_condensate}.
We diagonalize $H'_{\mathrm{MF}}$ and introduce the Slater determinant by single-particle states.
Then, we perform the variation for the expectation value $\langle H'_{\mathrm{MF}} \rangle$ with respect to $\lambda$ and $\Delta^{i}$ as
\begin{eqnarray}
\frac{\partial}{\partial \lambda} \langle H'_{\mathrm{MF}} \rangle &=& 0, \\
\frac{\partial}{\partial \Delta^{i}} \langle H'_{\mathrm{MF}} \rangle &=& 0,
\end{eqnarray}
and obtain $\lambda$ and $\Delta^{i}$.
The ground-state energy is given by substituting the obtained $\lambda$ and $\Delta^{i}$ into $\langle H'_{\mathrm{MF}} \rangle$.

In the followings, to demonstrate the mean-field calculation explicitly, we set $\epsilon_{k}=\epsilon$ as a simple case, where the diagonalization of $H'_{\mathrm{MF}}$ can be performed analytically.
Such simplification does not change the essence of the discussion.
By using $\{c_{k \sigma},f_{\sigma}\}$ ($k=1$, $\dots$, $N$, $\sigma = \uparrow$, $\downarrow$), we write the mean-field Hamiltonian $H'_{\mathrm{MF}}$, with the $2(N+1) \times 2(N+1)$ matrix ${\cal H}'_{cf}$,
\begin{widetext}
\begin{eqnarray}
{\cal H}'_{cf}
=
\left(
\begin{array}{cccc|cccccc}
 \epsilon & 0 & \cdots  & \Delta^{3\ast} & 0 & 0 &  \cdots & \Delta^{1\ast}\!-\!i\Delta^{2\ast} \\
 0 & \epsilon & \cdots  & \Delta^{3\ast} & 0 & 0 & \cdots & \Delta^{1\ast}\!-\!i\Delta^{2\ast} \\
 \vdots & \vdots  & \ddots  & \vdots & \vdots & \vdots &  \ddots & \vdots \\
 \Delta^{3} & \Delta^{3} & \cdots & \lambda & \Delta^{1\ast}\!-\!i\Delta^{2\ast} & \Delta^{1\ast}\!-\!i\Delta^{2\ast} & \cdots & 0 \\  
 \hline
 0 & 0 &  \cdots & \Delta^{1}\!+\!i\Delta^{2} & \epsilon & 0 & \cdots  & -\Delta^{3\ast} \\
 0 & 0 & \cdots & \Delta^{1}\!+\!i\Delta^{2} & 0 & \epsilon & \cdots  & -\Delta^{3\ast} \\
 \vdots & \vdots &  \ddots & \vdots & \vdots & \vdots  & \ddots  & \vdots \\
 \Delta^{1}\!+\!i\Delta^{2} & \Delta^{1}\!+\!i\Delta^{2} & \cdots & 0 &  -\Delta^{3} & -\Delta^{3} & \cdots & \lambda \\
\end{array}
\right),
\end{eqnarray}
\end{widetext}
as
\begin{eqnarray}
H'_{\mathrm{MF}} 
= 
\psi^{\dag} {\cal H}'_{cf}
\psi
-\frac{1}{g} \sum |\Delta^{i}|^{2} - \lambda,
\end{eqnarray}
with $\psi$ in Eq.~(\ref{eq:psi_field}).
It is worth to note that $g<0$ should be satisfied for the triplet condensate because the positivity of $-\sum |\Delta^{i}|^{2}/g$ supports the stability of the ground state in $H'_{\mathrm{MF}}$.
We diagonalize ${\cal H}_{cf}'$ analytically as
\begin{widetext}
\begin{eqnarray}
{\cal H}_{cf}'^{\mathrm{diag}}
&=&
\mathrm{diag}\left( \epsilon,\dots,\epsilon,\frac{1}{2} (\epsilon+\lambda-D'),\frac{1}{2} (\epsilon+\lambda+D'),\epsilon,\dots,\epsilon,\frac{1}{2} (\epsilon+\lambda-D'),\frac{1}{2} (\epsilon+\lambda+D') \right) \nonumber \\
&=&
\mathrm{diag}\left( E'_{1},\dots,E'_{N-1},E'_{N},E'_{N+1},E'_{1},\dots,E'_{N-1},E'_{N},E'_{N+1} \right),
\end{eqnarray}
\end{widetext}
with
\begin{eqnarray}
 D'=\sqrt{(\epsilon-\lambda)^{2}+4N\sum |\Delta^{i}|^{2}}.
\end{eqnarray}
Then, we obtain the mean-field Hamiltonian $H_{\mathrm{MF}}'$, as we discussed in Section~\ref{sec:isosinglet_condensate}.

Let us consider the system with one valence nucleon.
The extension to $n$ valence nucleons is straightforward, as discussed later.
We consider the isospin $\uparrow$, $\downarrow$ states with energy $E'_{N}$ as the most stable state.
For example, for the case of $\Delta^{1}=\Delta^{2}=0$, $\Delta^{3}\neq 0$, we define
\begin{eqnarray}
a_{0 \uparrow} &=& \frac{1}{\sqrt{2N}} \sqrt{1-\frac{\epsilon-\lambda}{D'}} \left( c_{1\uparrow} + \cdots + c_{N\uparrow} \right) \nonumber \\
&&- \frac{1}{\sqrt{2}} \sqrt{1+\frac{\epsilon-\lambda}{D'}} f_{\uparrow}, \\
a_{1 \uparrow} &=& \frac{1}{\sqrt{2N}} \sqrt{1+\frac{\epsilon-\lambda}{D'}} \left( c_{1\uparrow} + \cdots + c_{N\uparrow} \right) \nonumber \\
&&+ \frac{1}{\sqrt{2}} \sqrt{1-\frac{\epsilon-\lambda}{D'}} f_{\uparrow}, \\
a_{0 \downarrow} &=& \frac{1}{\sqrt{2N}} \sqrt{1-\frac{\epsilon-\lambda}{D'}} \left( c_{1\downarrow} + \cdots + c_{N\downarrow} \right) \nonumber \\
&&+ \frac{1}{\sqrt{2}} \sqrt{1+\frac{\epsilon-\lambda}{D'}} f_{\downarrow}, \\
a_{1 \downarrow} &=& \frac{1}{\sqrt{2N}} \sqrt{1+\frac{\epsilon-\lambda}{D'}} \left( c_{1\downarrow} + \cdots + c_{N\downarrow} \right) \nonumber \\
&&- \frac{1}{\sqrt{2}} \sqrt{1-\frac{\epsilon-\lambda}{D'}} f_{\downarrow}.
\end{eqnarray}
The ground state is given as $| \psi'_{0} \rangle = {a_{0\uparrow}}^{\!\dag} {a_{0\downarrow}}^{\!\dag} | 0 \rangle$.
In general cases, the mean-field energy $E'_{\mathrm{MF}}(\lambda,\{\Delta^{i}\})=\langle H'_{\mathrm{MF}} \rangle$ is represented as
\begin{eqnarray}
E'_{\mathrm{MF}}(\lambda,\{\Delta^{i}\}) = 2E'_{N} - \frac{1}{g} \sum |\Delta^{i}|^{2} - \lambda.
\end{eqnarray}
Performing the variation for $E'_{\mathrm{MF}}(\lambda,\{\Delta^{i}\})$ with respect to $\lambda$ and $\Delta^{i}$,
\begin{eqnarray}
 \frac{\partial}{\partial \lambda} E'_{\mathrm{MF}} =0, \hspace{1em}  \frac{\partial}{\partial \Delta^{i}} E'_{\mathrm{MF}} =0,
\end{eqnarray}
we obtain
\begin{eqnarray}
\lambda = \epsilon, \hspace{1em} \sqrt{ \sum |\Delta^{i}|^{2} } = -\sqrt{N}g.
\end{eqnarray}
We parametrize $\Delta^{i}$ by angles $\theta$, $\varphi$ as
\begin{eqnarray}
 \Delta^{1} &=& \Delta_{0} \sin \theta \cos \varphi, \\
 \Delta^{2} &=& \Delta_{0} \sin \theta \sin \varphi, \\
 \Delta^{3} &=& \Delta_{0} \cos \theta,
\end{eqnarray}
with $\Delta_{0} = -\sqrt{N}g$.
The ground-state energy $E'_{\mathrm{MF}}$ in the mean-field approximation is
\begin{eqnarray}
 E'_{\mathrm{MF}}(\epsilon,\Delta_{0},\theta,\varphi) = \epsilon + Ng.
\end{eqnarray}
We note that there is degeneracy for changing the angle parameter $(\theta,\varphi)$.
Therefore, the isospin symmetry $\mathrm{SU}(2) \simeq \mathrm{SO}(3)$ is broken to the $\mathrm{U}(1)$ symmetry in the ground state with the isotriplet condensate.
For example, in the case of $\Delta^{1}=\Delta^{2}=0$, $\Delta^{3}\neq 0$, the ground state has the $\mathrm{U}(1)$ symmetry in which the element is given by $e^{\alpha T_{3}} \in \mathrm{U}(1)$ for generator $T_{3}$ with $\alpha$ a real number parameter.
We note that there is no symmetry breaking of isospin in isosinglet condensate in Section~\ref{sec:isosinglet_condensate}.

The ground-state energy is given by including the energy shift $\langle \psi'_{0}| H_{\mathrm{fluc}} | \psi'_{0} \rangle=-Ng$, as
\begin{eqnarray}
E'_{\mathrm{MF}+\mathrm{shift}}
&=&\langle \psi'_{0}| H_{\mathrm{MF}} | \psi'_{0} \rangle+\langle \psi'_{0}| H_{\mathrm{fluc}} | \psi'_{0} \rangle \nonumber \\
&=&(\epsilon + Ng) + (-Ng) \nonumber \\
&=&\epsilon,
\label{eq:mean-field_energy_1}
\end{eqnarray}
which is quite different from the exact value $\epsilon + Ng/2$ in Section~\ref{sec:exact}.
Therefore, the mean-field approach does not give the good approximation for the isotriplet condensate with $g<0$.
The correction is given by RPA as it will be shown.

We leave a comment.
In the mean-field approximation, we obtain no bound state as shown in Eq.~(\ref{eq:mean-field_energy_1}) for $g<0$.
It may be worthwhile to compare this result with the behavior of the effective coupling strength of the Kondo interaction in the infrared limit \cite{0022-3719-3-12-008,Abrikosov1965} (see also Refs.~\cite{Hewson,Yamada}).
It is known that the effective coupling for $g<0$ becomes zero in the low-energy limit in the renormalization group method, when the coupling strength is renormalized by  including the loop effect dressed by particle-hole pairs near the Fermi surface.
This means that the interaction for $g<0$ vanishes in the low-energy limit and that no bound state is formed.
In the case of $g>0$, on the other hand, the effective coupling strength becomes large in the low-energy limit, leading to the formation of the bound state.
This is consistent with the existence of the bound state in the mean-field approximation for $g>0$ as shown in Eq.~(\ref{eq:mean-field_energy_0}).
In literature, the dependence of the existence/non-existence of the bound state on the sign of the coupling constant in the Kondo interaction was presented  for the first time in Ref.~\cite{Yosida:1966}.
It is interesting to see that the present analysis in the mean-field approximation gives the same result with the known results.

\subsubsection{Fluctuation effect ---RPA---}

We consider the fluctuation effect by RPA.
In the following, we consider the case of $\Delta^{1}=\Delta^{2}=0$, $\Delta^{3}=-\sqrt{N}g$ ($\theta=0$).
The other case can be discussed similarly.
The Hamiltonian (\ref{eq:H}) 
 is rewritten as
\begin{eqnarray}
H 
&=&
 \left( \epsilon + \frac{7}{4} Ng \right) ({a_{0\uparrow}}^{\!\dag} a_{0\uparrow} + {a_{0\downarrow}}^{\!\dag} a_{0\downarrow}) \nonumber \\
&& + \left( \epsilon + \frac{7}{4} Ng \right) ({a_{1\uparrow}}^{\!\dag} a_{1\uparrow} + {a_{1\downarrow}}^{\!\dag} a_{1\downarrow})
 \nonumber \\
&&
+Ng ( {a_{0\uparrow}}^{\!\dag} {a_{1\uparrow}}^{\!\dag} a_{0\uparrow} a_{1\uparrow} + {a_{0\downarrow}}^{\!\dag} {a_{1\downarrow}}^{\!\dag} a_{0\downarrow}a_{1\downarrow}) \nonumber \\
&&
+ \frac{1}{2} Ng ( {a_{0\uparrow}}^{\!\dag} - {a_{1\uparrow}}^{\!\dag} ) ( {a_{0\downarrow}}^{\!\dag} + {a_{1\downarrow}}^{\!\dag} ) ( a_{0\uparrow} - a_{1\uparrow} ) \nonumber \\
&& \hspace{2.1em} \times( a_{0\downarrow} + a_{1\downarrow} )
\nonumber \\
&&
+ \frac{1}{2} Ng ( {a_{0\uparrow}}^{\!\dag} + {a_{1\uparrow}}^{\!\dag} ) ( {a_{0\downarrow}}^{\!\dag} - {a_{1\downarrow}}^{\!\dag} ) ( a_{0\uparrow} + a_{1\uparrow} ) \nonumber \\
&& \hspace{2.1em} \times( a_{0\downarrow} - a_{1\downarrow} )
\nonumber \\
&&
+ \left( - \frac{1}{4} \right) Ng ( {a_{0\uparrow}}^{\!\dag} - {a_{1\uparrow}}^{\!\dag} ) ( {a_{0\downarrow}}^{\!\dag} + {a_{1\downarrow}}^{\!\dag} ) ( a_{0\uparrow} + a_{1\uparrow} ) \nonumber \\
&& \hspace{4.8em} \times ( a_{0\downarrow} - a_{1\downarrow} )
\nonumber \\
&&
+ \left( - \frac{1}{4} \right) Ng ( {a_{0\uparrow}}^{\!\dag} + {a_{1\uparrow}}^{\!\dag} ) ( {a_{0\downarrow}}^{\!\dag} - {a_{1\downarrow}}^{\!\dag} ) ( a_{0\uparrow} - a_{1\uparrow} ) \nonumber \\
&& \hspace{4.8em} \times( a_{0\downarrow} + a_{1\downarrow} )
\nonumber \\
&&
+ \left( - \frac{1}{4} \right) Ng ( {a_{0\uparrow}}^{\!\dag} a_{1\uparrow} + {a_{0\downarrow}}^{\!\dag} a_{1\downarrow} + {a_{1\uparrow}}^{\!\dag} a_{0\uparrow} + {a_{1\downarrow}}^{\!\dag} a_{0\downarrow} )
\nonumber \\
&&+\left(-\epsilon-3Ng\right)
+\sum_{k=1}^{N-1} \epsilon \, {d_{k \sigma}}^{\!\dag} d_{k\sigma}.
\label{eq:H_RPA_1}
\end{eqnarray}
Considering the fluctuation around the ground state $| \psi'_{0} \rangle$, we solve the RPA equation
\begin{eqnarray}
\left(
\begin{array}{cc}
 A & B  \\
 -B^{\ast} & -A^{\ast}
\end{array}
\right)
\left(
\begin{array}{c}
 X  \\
 Y
\end{array}
\right)
=
\Omega_{\nu}
\left(
\begin{array}{c}
 X  \\
 Y
\end{array}
\right),
\end{eqnarray}
with
\begin{eqnarray}
A_{\mu\nu\rho\sigma} &=& \langle \psi'_{0} | \left[ {a_{0\nu}}^{\!\dag} a_{1\mu}, \left[ H, {a_{1\rho}}^{\!\dag} a_{0\sigma} \right] \right]  | \psi'_{0} \rangle 
\nonumber \\
&=&
\frac{3}{2}Ng \left( \delta_{\mu \uparrow} \delta_{\nu \uparrow} \delta_{\rho \downarrow} \delta_{\sigma \downarrow} + \delta_{\mu \downarrow} \delta_{\nu \downarrow} \delta_{\rho \uparrow} \delta_{\sigma \uparrow} \right) \nonumber \\
&&+(-1)Ng \left( \delta_{\mu \uparrow} \delta_{\nu \uparrow} \delta_{\rho \uparrow} \delta_{\sigma \uparrow} + \delta_{\mu \downarrow} \delta_{\nu \downarrow} \delta_{\rho \downarrow} \delta_{\sigma \downarrow} \right) \nonumber \\
&&+\left( -\frac{1}{2} \right) Ng \left( \delta_{\mu \uparrow} \delta_{\nu \downarrow} \delta_{\rho \uparrow} \delta_{\sigma \downarrow} + \delta_{\mu \downarrow} \delta_{\nu \uparrow} \delta_{\rho \downarrow} \delta_{\sigma \uparrow} \right), \nonumber \\
\end{eqnarray}
and
\begin{eqnarray}
-B_{\mu\nu\rho\sigma} &=& \langle \psi'_{0} | \left[ {a_{0\nu}}^{\!\dag} a_{1\mu}, \left[ H, {a_{0\sigma}}^{\!\dag} a_{1\rho} \right] \right]  | \psi'_{0} \rangle \nonumber \\
&=& \left( -\frac{1}{2} \right) Ng \left( \delta_{\mu \uparrow} \delta_{\nu \uparrow} \delta_{\rho \downarrow} \delta_{\sigma \downarrow} - \delta_{\mu \uparrow} \delta_{\nu \downarrow} \delta_{\rho \downarrow} \delta_{\sigma \uparrow} \right. \nonumber \\
&& \hspace{2em} \left. - \delta_{\mu \downarrow} \delta_{\nu \uparrow} \delta_{\rho \uparrow} \delta_{\sigma \downarrow} + \delta_{\mu \downarrow} \delta_{\nu \downarrow} \delta_{\rho \uparrow} \delta_{\sigma \uparrow} \right),
\end{eqnarray}
The RPA energy eigenvalues are obtained as
\begin{eqnarray}
\left\{ \Omega_{\nu} \right\} &=& \left\{ \Omega_{\pm1},\Omega_{0},\Omega_{0},\Omega_{0} \right\} \nonumber \\
&=& \left\{\mp \sqrt{6} Ng,0,0,0\right\}.
\end{eqnarray}
We note the ordering of signs in $\Omega_{\pm1}=\mp Ng$ because of $g<0$.
In the three zero-energy modes ($\Omega_{0}=0$), one is due to the energy degeneracy of $| \psi'_{0} \rangle = {a_{0\uparrow}}^{\!\dag}{a_{0\downarrow}}^{\!\dag} | 0 \rangle$ and $| \psi'_{1} \rangle = {a_{1\uparrow}}^{\!\dag}{a_{1\downarrow}}^{\!\dag} | 0 \rangle$ in the first two terms of the Hamiltonian (\ref{eq:H_RPA_1}).
However, this degeneracy is special in the mean-field approximation, and hence should be a spurious one.
The other two are the Nambu-Goldstone modes in the coset space $\mathrm{SU}(2)/\mathrm{U}(1)$, because the isospin symmetry in the ground state is broken from $\mathrm{SU}(2)$ to $\mathrm{U}(1)$, where, for example, the $\mathrm{U}(1)$ symmetry is given by $e^{\alpha T_{3}} \in \mathrm{U}(1)$ in the case of $\Delta^{1}=\Delta^{2}=0$, $\Delta^{3} \neq 0$.

From the above result, we obtain the RPA correlation energy \cite{Ring-Schuck,Rowe:1968zza,Hagino:1999mu,Hagino:2000ba}
\begin{eqnarray}
\Delta E'_{\mathrm{RPA}}
&=&
\frac{1}{2} \sum_{\nu>0} \Omega_{\nu} - \frac{1}{2} \mathrm{Tr}A \nonumber \\
&=&
\frac{1}{2} (3-\sqrt{6}) Ng \nonumber \\
&\simeq& 0.275 Ng.
\end{eqnarray}
Therefore, the ground-state energy in the mean-field approximation and the RPA is given by
\begin{eqnarray}
E'_{\mathrm{MF}+\mathrm{shift}+\mathrm{RPA}}
&=&
E'_{\mathrm{MF}+\mathrm{shift}}+\Delta E'_{\mathrm{RPA}} \nonumber \\
&=&
\epsilon + \frac{1}{2} (3-\sqrt{6}) Ng \nonumber \\
&\simeq&
\epsilon + 0.275 Ng.
\end{eqnarray}
When there are $n$ valence nucleons ($n \le 2N$), one valence nucleon participates in coupling to the impurity, and $n-1$ valence nucleons do not.
Therefore, the ground-state energy is modified to
\begin{eqnarray}
E_{\mathrm{MF}+\mathrm{shift}+\mathrm{RPA}}(n)
&\simeq&
n \,\epsilon + 0.275 Ng.
\end{eqnarray}
The obtained binding energy is about 55\% to the exact one ($Ng/2$).
Thus, the fluctuation by RPA cannot be neglected to obtain the ground-state energy in the mean-field approximation.

\section{Discussion: Competition between Kondo effect and nucleon correlations}

So far, we have discussed the correlation between an impurity and a valence nucleon as the Kondo effect, and assumed no correlation between valence nucleons.
In realistic nuclei, however, there are several correlations in valence nucleons which are not necessarily negligible.
In this section, we briefly consider two types of correlation in valence nucleons, the isospin symmetry breaking and the nucleon pairings, and discuss how the Kondo effect is affected by them (see for example Ref.~\cite{Balatsky:2006} as a review in the condensed matter systems).

\subsection{Competition between Kondo effect and isospin breaking}

We consider the isospin symmetry breaking in the valence nucleons.
We set the valence nucleon energies $\epsilon_{\uparrow}$ and $\epsilon_{\downarrow}$ for $\uparrow$ and $\downarrow$ components of isospin, respectively, and modify the kinetic term of the valence nucleon (\ref{eq:H0}),
\begin{eqnarray}
H_{0} \rightarrow H_{0} = \sum \epsilon_{k\sigma} {c_{k\sigma}}^{\!\dag} c_{k\sigma},
\end{eqnarray}
to include the isospin breaking in the single-particle states.
In the following, we set $\epsilon_{k\sigma}=\epsilon_{\sigma}$ for simplicity.
The calculation procedure in the mean-field approximation is essentially the same as discussed in Section~\ref{sec:isosinglet_condensate}.
We introduce the isospin breaking in the mean-field Hamiltonian Eq.~(\ref{eq:Ham_MF_0}).
Instead of the matrix (\ref{eq:H_MF_0}),
we define the generalized matrix with $2(N+1) \times 2(N+1)$ dimensions
\begin{eqnarray}
\tilde{\cal H}_{cf}
=
\left(
\begin{array}{cccc|cccccc}
 \epsilon_{\uparrow} & 0 & \cdots  & \Delta^{\ast} & 0 & 0 &  \cdots & 0 \\
 0 & \epsilon_{\uparrow} & \cdots  & \Delta^{\ast} & 0 & 0 & \cdots & 0 \\
 \vdots & \vdots  & \ddots  & \vdots & \vdots & \vdots &  \ddots & \vdots \\
 \Delta & \Delta & \cdots & \lambda & 0 & 0 & \cdots & 0 \\  
 \hline
 0 & 0 &  \cdots & 0 & \epsilon_{\downarrow} & 0 & \cdots  & \Delta^{\ast} \\
 0 & 0 & \cdots & 0 & 0 & \epsilon_{\downarrow} & \cdots  & \Delta^{\ast} \\
 \vdots & \vdots &  \ddots & \vdots & \vdots & \vdots  & \ddots  & \vdots \\
 0 & 0 & \cdots & 0 &  \Delta & \Delta & \cdots & \lambda \\
\end{array}
\right),
\end{eqnarray}
and perform the diagonalization as
\begin{widetext}
\begin{eqnarray}
 \tilde{\cal H}_{cf}^{\mathrm{diag}}
&=&
\mathrm{diag} \left( \epsilon_{\uparrow}, \dots, \epsilon_{\uparrow}, \frac{1}{2} (\epsilon_{\uparrow}+\lambda-D_{\uparrow}), \frac{1}{2} (\epsilon_{\uparrow}+\lambda+D_{\uparrow}), \epsilon_{\downarrow}, \dots, \epsilon_{\downarrow}, \frac{1}{2} (\epsilon_{\downarrow}+\lambda-D_{\downarrow}), \frac{1}{2} (\epsilon_{\downarrow}+\lambda+D_{\downarrow}) \right) \nonumber \\
&=&
\mathrm{diag} \left( E_{1 \uparrow}, \dots, E_{N-1 \uparrow},  E_{N \uparrow}, E_{N+1 \uparrow}, E_{1 \downarrow}, \dots, E_{N-1 \downarrow}, E_{N \downarrow}, E_{N+1 \downarrow} \right),
\end{eqnarray}
\end{widetext}
with
\begin{eqnarray}
 D_{\sigma}=\sqrt{(\epsilon_{\sigma}-\lambda)^{2}+4N|\Delta|^{2}},
\end{eqnarray}
for $\sigma=\uparrow$, $\downarrow$.
Instead of the original fields $\{c_{k \sigma},f_{\sigma}\}$, we define the new fields 
\begin{eqnarray}
\tilde{d}_{1\sigma} &=& \frac{1}{\sqrt{2}} \left( c_{1\sigma}-c_{2\sigma} \right), \nonumber \\
\tilde{d}_{2\sigma} &=& \frac{1}{\sqrt{2}} \left( c_{1\sigma}-c_{3\sigma} \right), \nonumber \\
&\vdots& \nonumber \\
\tilde{d}_{N-1\sigma} &=& \frac{1}{\sqrt{2}} \left(c_{1\sigma}-c_{N\sigma}\right), \nonumber \\
\tilde{d}_{N\sigma} &=& \frac{1}{\sqrt{2N}} \sqrt{1-\frac{\epsilon_{\sigma}-\lambda}{D_{\sigma}}} \left( c_{1\sigma}+\dots+c_{N\sigma} \right) \nonumber \\
&& - \frac{1}{\sqrt{2}} \sqrt{1+\frac{\epsilon_{\sigma}-\lambda}{D_{\sigma}}} f_{\sigma}, \nonumber \\
\tilde{d}_{N+1\sigma} &=& \frac{1}{\sqrt{2N}} \sqrt{1-\frac{\epsilon_{\sigma}-\lambda}{D_{\sigma}}} \left( c_{1\sigma}+\dots+c_{N\sigma} \right) \nonumber \\
&&+ \frac{1}{\sqrt{2}} \sqrt{1+\frac{\epsilon_{\sigma}-\lambda}{D_{\sigma}}} f_{\sigma}, \nonumber
\end{eqnarray}
and rewrite the mean-field Hamiltonian (\ref{eq:Ham_MF_0}) as
\begin{eqnarray}
\tilde{H}_{\mathrm{MF}}
&=&
\phi^{\dag} \tilde{\cal H}_{cf}^{\mathrm{diag}}
\phi
+ \frac{|\Delta|^{2}}{g} - \lambda \nonumber \\
&=&
\sum E_{k \sigma} {\tilde{d}_{k \sigma}}^{\,\,\,\,\dag} \tilde{d}_{k \sigma} + \frac{|\Delta|^{2}}{g} - \lambda,
\end{eqnarray}
with $\phi$ in Eq.~(\ref{eq:phi_field}).
Supposing the energy of $\downarrow$ component is larger than that of $\uparrow$ component, we parametrize $\epsilon_{\uparrow}$ and $\epsilon_{\downarrow}$ by
 $\epsilon_{\uparrow} = \epsilon$ and 
 $\epsilon_{\downarrow} = \epsilon + \delta \epsilon$
with $\delta \epsilon > 0$.

For the system composed of one valence nucleon and an impurity,
we consider the ground state given by
\begin{eqnarray}
| \tilde{\psi}_{0} \rangle = {\tilde{d}_{N \uparrow}}^{\,\,\,\,\dag} {\tilde{d}_{N \downarrow}}^{\,\,\,\,\dag} | 0 \rangle.
\end{eqnarray}
For the ground-state energy,
\begin{eqnarray}
\tilde{E}_{\mathrm{MF}}
&=& \langle \tilde{\psi}_{0} | \tilde{H}_{\mathrm{MF}} | \tilde{\psi}_{0} \rangle \nonumber \\
&=& E_{N\uparrow} + E_{N\downarrow} + \frac{|\Delta|^{2}}{g} - \lambda,
\end{eqnarray}
we perform the variation with respect to $\lambda$ and $\Delta$,
\begin{eqnarray}
 \frac{\partial}{\partial \lambda} \tilde{E}_{\mathrm{MF}} =0, \hspace{1em}  \frac{\partial}{\partial \Delta} \tilde{E}_{\mathrm{MF}} =0,
\end{eqnarray}
and obtain $\lambda$ and $\Delta$ as
\begin{eqnarray}
\lambda = \epsilon+\frac{1}{2}\delta\epsilon, \hspace{1em} \Delta = \sqrt{N}g \sqrt{1-\frac{(\delta \epsilon)^{2}}{16N^{2}g^{2}}}.
\label{eq:lambda_Delta_breaking}
\end{eqnarray}
Then, the ground-state energy is given as
\begin{eqnarray}
 &&\tilde{E}_{\mathrm{MF}} \left( \epsilon+\frac{1}{2}\delta\epsilon,\sqrt{N}g \sqrt{1-\frac{(\delta \epsilon)^{2}}{16N^{2}g^{2}}} \right) \nonumber \\
 &=& \epsilon - Ng + \frac{1}{2} \delta \epsilon - \frac{(\delta \epsilon)^{2}}{16Ng}.
\end{eqnarray}
We note that the ground-state energy increases for non-zero $\delta \epsilon$ with $0< \delta \epsilon \le 4Ng$.
At the special value $\delta \epsilon = 4Ng$, we get $\tilde{E}_{\mathrm{MF}}=\epsilon$ and find no bound state.
According to the change of the ground-state energy by $\delta \epsilon$, we find that the strength of the gap function $|\Delta|$ in Eq.~(\ref{eq:lambda_Delta_breaking}) decreases and finally it becomes $|\Delta|=0$ at $\delta \epsilon = 4Ng$ where the bound state disappears.

For $\delta \epsilon > 4Ng$, on the other hand, the ground state is the isosinglet or isotriplet state given by $\tilde{d}_{k\sigma}^{\,\,\,\dag}$ ($k=1$, $\dots$, $N-1$) and $\tilde{d}_{N\sigma '}^{\,\,\,\dag}$ with the number of degeneracy factor $N-1$.
The ground-state energy is $\tilde{E}_{\mathrm{MF}}=\epsilon$ with $\lambda=\epsilon-Ng$ and $\Delta=0$.
The solution for $\delta \epsilon \le 4Ng$ is connected continuously to the solution for $\delta \epsilon > 4Ng$.

\subsection{Competition between Kondo effect and pairing of valence nucleons}

We consider the pairing interaction in valence nucleons, and discuss the competition between the Kondo effect and the pairing effect.
We consider two types of the pairing interaction: the isovector-type ($I=1$) pairing and the isoscalar-type ($I=0$) pairing.

We consider the isovector-type pairing interaction
\begin{widetext}
\begin{eqnarray}
 H_{\mathrm{pair}}^{1} 
&\!=\!&  -G_{1}
 \!\!\!\! \sum_{i,j:\mathrm{odd}}
 \!\! \left\{ {c_{i\uparrow}}^{\!\dag} {c_{i+1\uparrow}}^{\!\dag} c_{j+1\uparrow} c_{j\uparrow}
      \!+\! \frac{1}{2} \left( {c_{i\uparrow}}^{\!\dag} {c_{i+1 \downarrow}}^{\!\dag} \!+\! {c_{i\downarrow}}^{\!\dag} {c_{i+1 \uparrow}}^{\!\dag} \right)
                          \left( c_{j+1\uparrow} c_{j\downarrow} \!+\! c_{j+1\downarrow} c_{j\uparrow} \right)
      \!+\! {c_{i\downarrow}}^{\!\dag} {c_{i+1\downarrow}}^{\!\dag} c_{j+1\downarrow} c_{j\downarrow}
 \right\},
\end{eqnarray}
\end{widetext}
with the coupling constant $G_{1}>0$, where we give the pairing between the $i$th and $i+1$th single-particle states with $i$, $j$ being odd numbers, as the simple pairing model.
Because of the commutation relation
\begin{eqnarray}
 \left[ H_{\mathrm{pair}}^{1} , H_{\mathrm{K}} \right] \neq 0,
 \label{eq:Kondo_pairing_1}
\end{eqnarray}
for the Kondo interaction $H_{\mathrm{K}}$ (Eq.~(\ref{eq:HK})),
we find that 
the bound state by the Kondo interaction is affected by the isovector-type pairing.

The situation is different for the isoscalar-type pairing.
We consider the isoscalar-type pairing interaction
\begin{eqnarray}
 H_{\mathrm{pair}}^{0} &=& \frac{G_{0}}{2} \sum_{i,j:\mathrm{odd}}
\left( {c_{i\uparrow}}^{\!\dag} {c_{i+1 \downarrow}}^{\!\dag} - {c_{i\downarrow}}^{\!\dag} {c_{i+1 \uparrow}}^{\!\dag} \right) \nonumber \\
&& \hspace{3.5em} \times                         \left( c_{j+1\uparrow} c_{j\downarrow} - c_{j+1\downarrow} c_{j\uparrow} \right),
\end{eqnarray}
with the coupling constant $G_{0}/2>0$, where we give 
the pairing again for the $i$th and $i+1$th single-particle states with $i$, $j$ being odd numbers.
In this case, the commutation relation with the Kondo interaction $H_{\mathrm{K}}$ (Eq.~(\ref{eq:HK})) is given by
\begin{eqnarray}
 \left[ H_{\mathrm{pair}}^{0} , H_{\mathrm{K}} \right] = 0.
 \label{eq:Kondo_pairing_0}
\end{eqnarray}
Therefore, 
we find that the bound state by the Kondo interaction is not suffered from the isoscalar-type pairing.

Let us consider the eigenstates of the Hamiltonian $H_{0}+H_{\mathrm{K}}+H_{\mathrm{pair}}^{0,1}$.
We represent $|\psi_{\mathrm{K}}\rangle$ for the eigenstate of the Hamiltonian $H_{0}+H_{\mathrm{K}}$ (Eq.~(\ref{eq:H})) and $|\psi_{\mathrm{pair}}^{1}\rangle$ for the eigenstate of $H_{0}+H_{\mathrm{pair}}^{1}$.
We consider the case of $\epsilon_{k}=\epsilon$ in Eq.~(\ref{eq:H0}).
Because $H_{0}$, $H_{\mathrm{pair}}^{1}$ and $H_{\mathrm{K}}$ are non-commutative for each other (cf.~Eq.~(\ref{eq:Kondo_pairing_1})),
the eigenstate for $H_{0}+H_{\mathrm{K}}+H_{\mathrm{pair}}^{1}$ including both the Kondo effect and the isovector-type pairing is given by the sum of the tensor product of several states: $\sum_{i,j} \gamma_{ij} |\psi_{\mathrm{K}\,i}\rangle \otimes |\psi_{\mathrm{pair}\,j}^{1} \rangle$ with $|\psi_{\mathrm{K}\,i}\rangle$ and $|\psi_{\mathrm{pair}\,j}^{1} \rangle$ being the $i$th and $j$th states of $|\psi_{\mathrm{K}}\rangle$ and $|\psi_{\mathrm{pair}}^{1}\rangle$, respectively, and  $\gamma_{ij}$ being appropriate coefficients.
Therefore, $|\psi_{\mathrm{K}}\rangle$ and $|\psi_{\mathrm{pair}}^{1}\rangle$ are entangled by the Hamiltonian $H_{0}+H_{\mathrm{K}}+H_{\mathrm{pair}}^{1}$.
On the other hand, in the case of the isoscalar-type pairing, because $H_{0}$, $H_{\mathrm{pair}}^{1}$ and $H_{\mathrm{K}}$ are commutative (cf.~Eq.~(\ref{eq:Kondo_pairing_0})), the eigenstate of the Hamiltonian $H_{0}+H_{\mathrm{K}}+H_{\mathrm{pair}}^{0}$ is given by a simple tensor product:  $|\psi_{\mathrm{K}}\rangle \otimes |\psi_{\mathrm{pair}}^{0}\rangle$, where $|\psi_{\mathrm{pair}}^{0}\rangle$ is the eigenstate of $H_{0}+H_{\mathrm{pair}}^{1}$.
Therefore, $|\psi_{\mathrm{K}}\rangle$ and $|\psi_{\mathrm{pair}}^{0}\rangle$ are disentangled by the Hamiltonian $H_{0}+H_{\mathrm{K}}+H_{\mathrm{pair}}^{0}$.

We leave a comment for the case that the interaction between an impurity and a valence nucleon has no isospin-exchange, in contrast to the Kondo interaction (\ref{eq:HK}).
For the interaction with no isospin-exchange
 $H_{\mathrm{NK}} = g' \sum {c_{i\sigma}}^{\!\dag} c_{j\sigma}$
with the coupling constant $g'$,
we find that the commutations with the pairing interactions are given by $\left[ H_{\mathrm{pair}}^{1} , H_{\mathrm{NK}} \right] \neq 0$ and $\left[ H_{\mathrm{pair}}^{0} , H_{\mathrm{NK}} \right] \neq 0$.
Therefore, the eigenstate of $H_{0}+H_{\mathrm{NK}}$, $|\psi_{\mathrm{NK}}\rangle$, is affected both by the isoscalar-type pairing and by the isovector-type pairing, and hence $|\psi_{\mathrm{NK}}\rangle$ becomes entangled with $|\psi_{\mathrm{pair}}^{0,1}\rangle$ by the Hamiltonian $H_{0}+H_{\mathrm{NK}}+H_{\mathrm{pair}}^{0,1}$.

The above properties of the entanglement are obtained as the exact solutions of the Hamiltonians.
Because they should be maintained also for the approximate solutions in the mean-field approach,
it gives us a guidance to check the validity of the mean-field approximation.
Further detailed analysis will be left for future studies.

\section{Conclusion}

We consider the Kondo effect in charm/bottom nuclei with a $\bar{D}$, $B$ meson bound as a heavy impurity, and discuss the binding energy for the isospin-exchange interaction between an impurity and a valence nucleon.
We consider the discrete energy-levels of valence nucleons in the charm/bottom nuclei.
By  introducing the auxiliary fermion field for the isospin of the $\bar{D}$, $B$ meson and extending the Fock space to include the multiple numbers of the impurity, we perform the mean-field approximation and the RPA calculation.
Based on the simple model which is analytically solvable,
we present that the approximate energy is comparable with the exact one, and find that the mean-field approach is valid as the ground state with the Kondo effect, when the fluctuation effect is included by the RPA.
The approach by the man-field approximation and the RPA is applicable to general cases, for example, where the interaction between an impurity and a valence nucleon is given by more realistic form and the structure of discrete energy-levels for valence nucleons are more complex.

In the present discussion, we consider a $\bar{D}$, $B$ meson ($I=1/2$) as a heavy impurity in charm/bottom nuclei.
We may also consider the case of a $\Sigma_{c}$, $\Sigma_{b}$ baryon with isospin $I=1$ \cite{Liu:2011xc,Maeda:2015hxa}.
In this case, the isospin-exchange interaction between a $\Sigma_{c}$, $\Sigma_{b}$ baryon and a valence nucleon will induce the bound state between them.
However, due to the isospin $I=1$ of the $\Sigma_{c}$, $\Sigma_{b}$ baryon, this bound state has still a finite isospin, $I=1/2$, and hence it can attract another nucleon to form the isosinglet bound state.
This is a three-body bound state composed of a $\Sigma_{c}$, $\Sigma_{b}$ baryon and two valence nucleons.
Because the two valence nucleons have isospin one as a subsystem, we expect that the properties of this three-body bound state can be affected by the isovector-type pairing for valence nucleons, rather than by the isoscalar-type pairing.
In any case, to study the Kondo effect in charm/bottom nuclei in various situations will be important for both experimental and theoretical researches.

\section*{Acknowledgments}
The author thanks M.~Oka and K.~Ohtani for fruitful discussions and useful comments. This work is supported by the Grant-in-Aid for Scientific Research (Grant No.~25247036 and No.~15K17641) from Japan Society for the Promotion of Science (JSPS).

\appendix

\section{Including complete fluctuation effect for isosinglet condensate ($g>0$)}
\label{sec:complete_fluctuation}

We include the fluctuation effect completely beyond the RPA for the Hamiltonian (\ref{eq:H}) for $g>0$ (see also Eq.~(\ref{eq:H_RPA_0})).
In general form of Hamiltonian, such procedure is not necessarily always possible.
However, the present simple model enables us to obtain the exact solution by complete diagonalization.

We suppose the ground state wave function as
\begin{eqnarray}
 |\psi\rangle &=& c_{00} \, {a_{0\uparrow}}^{\!\dag}{a_{0\downarrow}}^{\!\dag}|0\rangle + c_{01} \, \left({a_{0\uparrow}}^{\!\dag}{a_{1\downarrow}}^{\!\dag}-{a_{0\downarrow}}^{\!\dag}{a_{1\uparrow}}^{\!\dag} \right) |0\rangle \nonumber \\
 && + c_{11} \, {a_{1\uparrow}}^{\!\dag}{a_{1\downarrow}}^{\!\dag}|0\rangle,
\end{eqnarray}
with three isosinglet bases ${a_{0\uparrow}}^{\!\dag}{a_{0\downarrow}}^{\!\dag}|0\rangle$, $({a_{0\uparrow}}^{\!\dag}{a_{1\downarrow}}^{\!\dag} \!-\!{a_{0\downarrow}}^{\!\dag}{a_{1\uparrow}}^{\!\dag})|0\rangle$, ${a_{1\uparrow}}^{\!\dag}{a_{1\downarrow}}^{\!\dag}|0\rangle$.
We obtain the energy eigenvalue $E$ and the coefficients $c_{00}$, $c_{01}$, $c_{11}$ by solving the eigenvalue equation $H|\psi\rangle = E|\psi\rangle$;
\begin{eqnarray}
\hspace*{-2.2em}
\left(
\begin{array}{ccc}
 \epsilon \!-\! Ng & \frac{1}{2}Ng & \frac{1}{2}Ng \\
 \frac{1}{4}Ng & \epsilon \!-\! \frac{1}{2}Ng  & \frac{1}{4}Ng  \\
 \frac{1}{2}Ng & \frac{1}{2}Ng  & \epsilon \!-\! Ng  
\end{array}
\right)
\!\!\!
\left(
\begin{array}{c}
 c_{00} \\
 c_{01} \\
 c_{11} 
\end{array}
\right)
\!=\!
E
\!
\left(
\begin{array}{c}
 c_{00} \\
 c_{01} \\
 c_{11} 
\end{array}
\right).
\end{eqnarray}
The solutions are
\begin{eqnarray}
E^{(0)} &=& \epsilon - \frac{3}{2} Ng, \\
E^{(1)} &=& \epsilon - Ng, \\
E^{(2)} &=& \epsilon,
\end{eqnarray}
with the coefficients
\begin{eqnarray}
\left(
\begin{array}{c}
 c_{00}^{(0)} \\
 c_{01}^{(0)} \\
 c_{11}^{(0)} 
\end{array}
\right)
&=&
\left(
\begin{array}{c}
 -1 \\
 0 \\
 1
\end{array}
\right), \\
\left(
\begin{array}{c}
 c_{00}^{(1)} \\
 c_{01}^{(1)} \\
 c_{11}^{(1)} 
\end{array}
\right)
&=&
\left(
\begin{array}{c}
 1 \\
 -1 \\
 1 
\end{array}
\right), \\
\left(
\begin{array}{c}
 c_{00}^{(2)} \\
 c_{01}^{(2)} \\
 c_{11}^{(2)} 
\end{array}
\right)
&=&
\left(
\begin{array}{c}
 1 \\
 1 \\
 1 
\end{array}
\right),
\end{eqnarray}
and the corresponding wave functions
\begin{widetext}
\begin{eqnarray}
| \psi^{(0)} \rangle &=& \frac{1}{\sqrt{2N}} \left( ({c_{1\uparrow}}^{\!\dag}+\dots+{c_{N\uparrow}}^{\!\dag}){f_{\downarrow}}^{\!\dag}  - ({c_{1\downarrow}}^{\!\dag}+\dots+{c_{N\downarrow}}^{\!\dag}){f_{\uparrow}}^{\!\dag} \right) | 0 \rangle, \label{eq:psi0} \\ 
| \psi^{(1)} \rangle &=& {f_{\uparrow}}^{\!\dag} {f_{\downarrow}}^{\!\dag} | 0 \rangle, \label{eq:psi1} \\
| \psi^{(2)} \rangle &=& \frac{1}{N} ({c_{1\uparrow}}^{\!\dag}+\dots+{c_{N\uparrow}}^{\!\dag}) ({c_{1\downarrow}}^{\!\dag}+\dots+{c_{N\downarrow}}^{\!\dag}) | 0 \rangle, \label{eq:psi2}
\end{eqnarray}
\end{widetext}
for respective eigenvalues.
We note that, for $n$ valence nucleons ($n \le 2N$), one of $n$ valence nucleons participates in the coupling to the impurity and the left $n-1$ valence nucleons do no, as discussed in the text.
Therefore, for $n$ valence nucleons, the energies of the Hamiltonian (\ref{eq:H}) are
\begin{eqnarray}
E^{(0)}(n) &=& n \, \epsilon - \frac{3}{2} Ng, \\
E^{(1)}(n) &=& n \, \epsilon - Ng, \\
E^{(2)}(n) &=& n \, \epsilon.
\end{eqnarray}
Here we remember that the Fock space is extended to multiple numbers of the impurity in the mean-field approach.
In the three states $| \psi^{(0)} \rangle$, $| \psi^{(1)} \rangle$ and $| \psi^{(2)} \rangle$, only the ground state $| \psi^{(0)} \rangle$ gives the exact solution as discussed in Section~\ref{sec:exact}.
The other two excited states $| \psi^{(1)} \rangle$ and $| \psi^{(2)} \rangle$ are the spurious states, because the number of the auxiliary fermions by $f_{\sigma}$ are 0 and 2, respectively, and hence they should be discarded as the real solution of the Hamiltonian (\ref{eq:H}).

The analogous discussion will be applied to the case of $g<0$ with the isotriplet condensate.

\end{document}